\documentclass[11pt]{article}

\usepackage[left=2cm, right=2cm, top=2.5cm, bottom=2.5cm]{geometry}
\usepackage{bm}

\usepackage[x11names]{xcolor}
\usepackage{fancyhdr, amssymb, cancel, amsmath, graphicx, pgfplots, tikz}
\usepackage{float}

\newcommand{\stylecolor}{black}

\usepackage[colorlinks=true, urlcolor=violet, linkcolor=blue, citecolor=red, hyperindex=true, linktocpage=true]{hyperref}

\pgfplotsset{samples=200}

\usepackage[explicit]{titlesec}

\newcommand*\sectionlabel{}
\titleformat{\section}
  {\gdef\sectionlabel{}
   \Large\bfseries\scshape}
  {\gdef\sectionlabel{\thesection. }}{0pt}
  {\begin{tikzpicture}[remember picture,overlay]
	\draw (-0.2, 0) node[right] {\textsf{\sectionlabel#1}};
	\draw[thick] (0, -0.4) -- (\textwidth, -0.4);
       \end{tikzpicture}
  }
\titlespacing*{\section}{0pt}{15pt}{20pt}

\newcommand*\subsectionlabel{}
\titleformat{\subsection}
  {\gdef\subsectionlabel{}
   \large\bfseries\scshape}
  {\gdef\subsectionlabel{\thesubsection.\ \  }}{0pt}
  {\begin{tikzpicture}[remember picture,overlay]
    	\draw (-0.15, 0) node[right] {\textsf{\subsectionlabel#1}};
       \end{tikzpicture}
  }
\titlespacing*{\subsection}{0pt}{10pt}{10pt}

\newcommand*\subsubsectionlabel{}
\titleformat{\subsubsection}
  {\gdef\subsubsectionlabel{}
   \bfseries\scshape}
  {\gdef\subsubsectionlabel{\thesubsubsection.\ \  }}{0pt}
  {\begin{tikzpicture}[remember picture,overlay]
    	\draw (-0.15, 0) node[right] {\textsf{\subsubsectionlabel#1}};
       \end{tikzpicture}
  }
\titlespacing*{\subsubsection}{0pt}{7pt}{7pt}

\pgfplotsset{every axis legend/.append style={at={(1.02,1)},anchor=north west}}

\newcommand{\titletext}{Conformal field theories in a periodic potential:
results from holography and field theory}

\newcommand{\beq}{\begin{equation}}
\newcommand{\eeq}{\end{equation}}
\newcommand{\ba}{\begin{array}{ccc}}
\newcommand{\ea}{\end{array}}
\newcommand{\nn}{\nonumber \\}
\newcommand{\bq}{{\bm q}}

\begin{document}
\thispagestyle{empty}

\begin{equation*}
\begin{tikzpicture}
\draw (0.5\textwidth, -3) node[text width = \textwidth] {{\huge \begin{center} \color{\stylecolor} \textsf{\textbf{Conformal field theories in a periodic potential:\\
results from holography and field theory}} \end{center}}}; 
\end{tikzpicture}
\end{equation*}
\begin{equation*}
\begin{tikzpicture}
\draw (0.5\textwidth, 0.1) node[text width=\textwidth] {\large \color{black} $\text{\textsf{Paul Chesler$^{a,b}$, Andrew Lucas$^b$ and Subir Sachdev$^b$}}$};
\draw (0.5\textwidth, -0.5) node[text width=\textwidth] {\small  $^a$\textsf{Center for Theoretical Physics, Massachusetts Institute of Technology, Cambridge, MA 02139, USA}};
\draw (0.5\textwidth, -1) node[text width=\textwidth] {\small  $^b$\textsf{Department of Physics, Harvard University, Cambridge, MA 02138, USA}};
\end{tikzpicture}
\end{equation*}
\begin{equation*}
\begin{tikzpicture}
\draw (0.5\textwidth, -6) node[below, text width=0.8\textwidth] {\small  
We study 2+1 dimensional conformal field theories (CFTs) with a globally conserved U(1) charge,
placed in a chemical potential which is periodically modulated along the spatial direction $x$ with zero average:
$\mu (x) = V \cos (k x)$. The dynamics of such theories depends only on the dimensionless ratio $V/k$,
and we expect that they flow in the infrared to new CFTs whose universality class changes as a function of $V/k$.
We compute the frequency-dependent conductivity of strongly-coupled CFTs using holography of the 
Einstein-Maxwell theory in 4-dimensional anti-de Sitter space. We compare the results with the corresponding computation
of weakly-coupled CFTs, perturbed away from the CFT of free, massless 
Dirac fermions (which describes graphene at low energies). We find that the results of the two computations have 
significant qualitative similarities. However, differences do appear in the vicinities of an infinite discrete set of values of $V/k$:
the universality class of the infrared CFT changes at these values in the weakly-coupled theory, by the emergence of new zero modes of Dirac fermions which are remnants of local Fermi surfaces.
The infrared theory changes continuously in holography, and 
the classical gravitational theory does not capture the physics of the discrete transition points between the infrared CFTs.
We briefly note implications for a non-zero average chemical potential.
};
\end{tikzpicture}
\end{equation*}
\begin{equation*}
\begin{tikzpicture}
\draw (0, -13.1) node[right, text width=0.5\textwidth] {\texttt{pchesler@physics.harvard.edu\\ lucas@fas.harvard.edu \\ sachdev@g.harvard.edu}};
\draw (\textwidth, -13.1) node[left] {\textsf{\today}};
\end{tikzpicture}
\end{equation*}

\tableofcontents

\pagestyle{fancy}
\renewcommand{\headrulewidth}{0pt}
\fancyhead{}

\fancyhead[L] {\textsf{\titletext}}
\fancyhead[R] {\textsf{\thepage}}
\fancyfoot{}

\section{Introduction}
\label{sec:intro}

A powerful method of studying non-Fermi liquid metallic systems is to apply a chemical potential to a strongly-coupled conformal field
theory (CFT) with a globally conserved U(1) charge; this allows use of the AdS/CFT correspondence by applying the chemical potential to the dual gravitational theory \cite{nernst,gubser,hhh,sscomp}. The simplest Einstein-Maxwell theory on AdS$_4$ already captures many key features of a 2+1 dimensional correlated metal at non-zero temperatures \cite{nernst}, although exposing its full low temperature Fermi surface structure will likely require quantum corrections from monopole operators \cite{faulkner,sscomp}. 

In studying the charge transport properties of a correlated metal, one immediately finds that the d.c. conductivity is infinite \cite{nernst}, 
a consequence of momentum conservation in the continuum theory: essentially all charge-current carrying states also have a non-zero momentum 
which is conserved, and so the current cannot decay. As some of the most striking experimental signatures of non-Fermi liquids
are in the d.c. conductivity,
it is important to add perturbations to the holographic theory which allow momentum fluctuations to relax to zero; such perturbations are invariably present in all condensed matter systems.
One approach is to add a dilute random concentration of impurities \cite{nernst,hh1,hh2}: this is useful at non-zero temperatures, 
but one faces the very difficult problem of understanding disordered non-Fermi liquids at zero temperature. 
Another recent idea has been to include a bulk graviton mass \cite{vegh,davison}, but the physical interpretation of this is not clear from the perspective
of the boundary field theory.

The present paper will focus on the idea of applying a periodic potential on the continuum 
boundary field theory \cite{flauger,hutasoit1,hh2,gary1,hutasoit2,gary2,gary3,schalm,ling}. This has the advantages
of being physically transparent and leaves open the possibility of understanding the true infrared (IR) behavior of the system.
Very interesting numerical studies of the influence of such potentials on bulk Einstein-Maxwell theories have been carried out recently
by Horowitz {\em et al.\/} \cite{gary1,gary2,gary3} and Liu {\em et al.} \cite{schalm}. 
These papers refer to the periodic potential as a ``lattice'', but we will not do so. 
We believe the term ``lattice'' should be limited to cases where there is a commensurability relation between the period of the lattice
and the density of matter, so that there are integer numbers of particles per unit cell. Such lattices do arise in holographic studies \cite{kachru1,kachru2,sscomp,kachru3,bao1},
and are linked to the condensation of monopole fields carrying dual magnetic charges \cite{sscomp}. We will not consider such lattices here,
and will be able to freely choose the period of the externally applied periodic potential.

Horowitz {\em et al.\/} \cite{gary1,gary2,gary3} studied systems with a periodic chemical potential of the form
\beq
\mu (x) = \mu_0 + V \cos (k x) \label{chemx}
\eeq
where $x$ is one of the spatial directions. Their average chemical potential was non-zero, $\mu_0 \neq 0$, and 
their d.c. conductivity remains infinite at zero temperature ($T$) even though translational symmetry is broken \cite{nernst,hh1,hh2}. 
So they had to turn on a non-zero $T$ to obtain solutions with a finite d.c.
conductivity. Consequently, their system was characterized by 4 energy scales: $\mu_0$, $V$, $k$, and $T$. There are 3 dimensionless ratios
one can make from these energies, and each ratio can be taken independently large or small. This makes identification of the true asymptotic regimes of various observables a very challenging task. They obtained some numerical evidence for non-Fermi liquid scalings, but numerous crossovers have made it difficult to precisely define the scaling regimes and to make analytic progress.

This paper will focus on the case 
\beq
\mu_0 = 0,
\eeq
where the average chemical potential vanishes (but we will note implications for $\mu_0 \neq 0$ in Section~\ref{sec:conc}). As we will argue below, in this case we expect a flow from the underlying ultraviolet (UV) 
CFT to an IR CFT,
and the d.c. conductivity is finite at $T=0$. Consequently, we are also able to examine the limit $T \rightarrow 0$.
The resulting system is now characterized by only 2 remaining energy scales, $V$ and $k$, and all physical properties are a function
only of the single dimensionless ratio $V/k$. It is our purpose here to describe this universal crossover as a function of $V/k$.

It is useful to first discuss this crossover in the context of boundary theory alone, without using holography. 
As a paradigm, we will consider in Section~\ref{sec:graphene} a theory of $N_f$ Dirac fermions $\psi_\alpha$ coupled to a SU($N_c$) gauge field $a_\mu$
in the limit of large $N_f$, as studied {\em e.g.\/} in Ref.~\cite{igor}, with the 2+1 dimensional Lagrangian\beq
\mathcal{L}_\psi = \sum_{\alpha=1}^{N_f} \overline{\psi}_\alpha \gamma^\mu (\partial_\mu - \mathrm{i} a_\mu) \psi_\alpha  
\label{L0}
\eeq
Note that these fermions carry gauge charges, and so in the holographic context they correspond to the `hidden' fermions of 
Refs.~\cite{liza,ogawa,hss},
and {\em not\/} the gauge-neutral `bulk' fermions of Refs.~\cite{sslee0,liu1,zaanen1,sean3,hong3,ssfl,schalm}.
At $N_f = \infty$, the gauge field can be neglected, and then this is a model for the low energy theory of graphene \cite{graphene1}. 
The chemical potential in Eq.~(\ref{chemx}) couples to the globally conserved U(1) charge density $\mathrm{i} \overline{\psi} \gamma^0 \psi$.
Our UV CFT is therefore described by $\mathcal{L}_\psi$, and we are interested in the IR physics in the presence of a non-zero $V$. 
In perturbation theory, we see that each action of $V$ transfers momentum $k$; all excitations of the UV CFT with momentum $k$
have an energy $\omega > k$ \cite{ssbook}, and so the perturbative expansion has positive energy denominators. 
In the $N_f = \infty$ theory, the spectrum of the fermionic excitations can be computed explicitly using the standard methods of solid
state physics, and for small $V/k$ we find that the low energy massless Dirac spectrum is preserved. However, these IR Dirac fermions
are anisotropic in space {\em i.e.\/} their velocities are different along the $x$ and $y$ directions. But this anisotropy is easily scaled away,
and so we conclude that for small $V/k$ the IR CFT is identical to the UV CFT. 

The situation changes dramatically for larger $V/k$, as will be described in Section~\ref{sec:graphene}. As has been studied in the graphene
literature \cite{park1,ho1,park2,park3,brey1,barbier1,wang1,brey2}, and even observed in experiment \cite{exp}, new Dirac zero modes emerge at certain non-zero momenta along the $y$ direction. 
We will argue in Section~\ref{sec:graphene} that these emergent Dirac points are {\em remnants of the local Fermi surfaces\/} that appear
in the limit of small $k$, where we can locally regard the chemical potential as constant (in a Born-Oppenheimer picture).
In general there are $N_D$ Dirac zeros for a given $V/k$, where $N_D$ 
increases monotonically in piecewise constant steps (and $N_D \rightarrow \infty$ as $V/k \rightarrow \infty$). At finite $N_f$, we have 
$N_D N_f$ Dirac fermions interacting with the gauge field $a_\mu$. Each of the $N_D$ nodes has its own anisotropic velocity, and so this anisotropy cannot be scaled by away by a change of co-ordinates. Nevertheless, we expect that under RG induced by the $a_\mu$ exchange, these velocities will flow to a common velocity \cite{hermele}, and the ultimate IR theory will be a CFT described by Eq.~(\ref{L0}) but now with $N_D N_f$ Dirac fermions. So there are an infinite set of possible IR CFTs, accessed with increasing $V/k$. The discrete transition points between 
these IR CFTs are described by separate low energy critical theories which are not CFTs.

Incidentally, the UV to IR flow described above appears to badly violate  
`$c$-theorems' because for $N_D >1$ there are many more low energy degrees of freedom in the IR than the UV.
However this is permitted
because our model breaks Lorentz invariance at 
intermediate scales \cite{swingle}.

Now let us consider the same physics for a strong-coupled CFT, as described by holography, as will be discussed in Section~\ref{sec:holo}. 
We use the simplest possible gravitational
model of a CFT with a conserved U(1) charge, the Einstein-Maxwell theory in 3+1 dimensions
\beq
\mathcal{L} = \frac{R-2\Lambda}{2\kappa^2} - \frac{1}{4e^2}F_{\mu\nu}F^{\mu\nu} \label{EM}
\eeq
where $\Lambda=-3/L^2$ with $L$ with AdS$_4$ radius of the UV theory, 
$F=\mathrm{d}A$ is the U(1) field strength, and $R$ is the Ricci scalar. 
Note that here $A$ is the gauge field dual to the conserved boundary U(1) charge, and is unrelated to $a_\mu$ above.
We will solve for the IR geometry of this theory in the presence of the boundary chemical potential Eq. 
(\ref{chemx}), by perturbative analytic methods for small $V/k$, and numerically for large $V/k$.  We find perturbatively that the IR geometry is a rescaled AdS$_4$, with the rescaling factors varying continuously as a function of $V/k$;
the resulting IR theory is a CFT, but with relative changes in the length scales associated to space and time. 

We computed the frequency-dependent conductivity, $\sigma (\omega)$, for charge transport along the $x$ direction, in both approaches.
In the absence of a potential, we have $\sigma (\omega) = \sigma_\infty$, a frequency-independent constant, which is a property of all CFTs
in 2+1 dimensions. The notation $\sigma_\infty$ refers to the fact that $\sigma (\omega \gg T) = \sigma_\infty$ at non-zero $T$, 
and we will limit ourselves to the $T \rightarrow 0$ limit in the present paper. The Einstein-Maxwell holographic theory
actually has $\sigma (\omega) = \sigma_\infty$ at all $T$ in the absence of a periodic potential, and this is due to a particle-vortex
self-duality \cite{m2cft,myers}. This self-duality is broken by the periodic potential.

\begin{figure}[H]
\begin{center}
\includegraphics[width=4.5in]{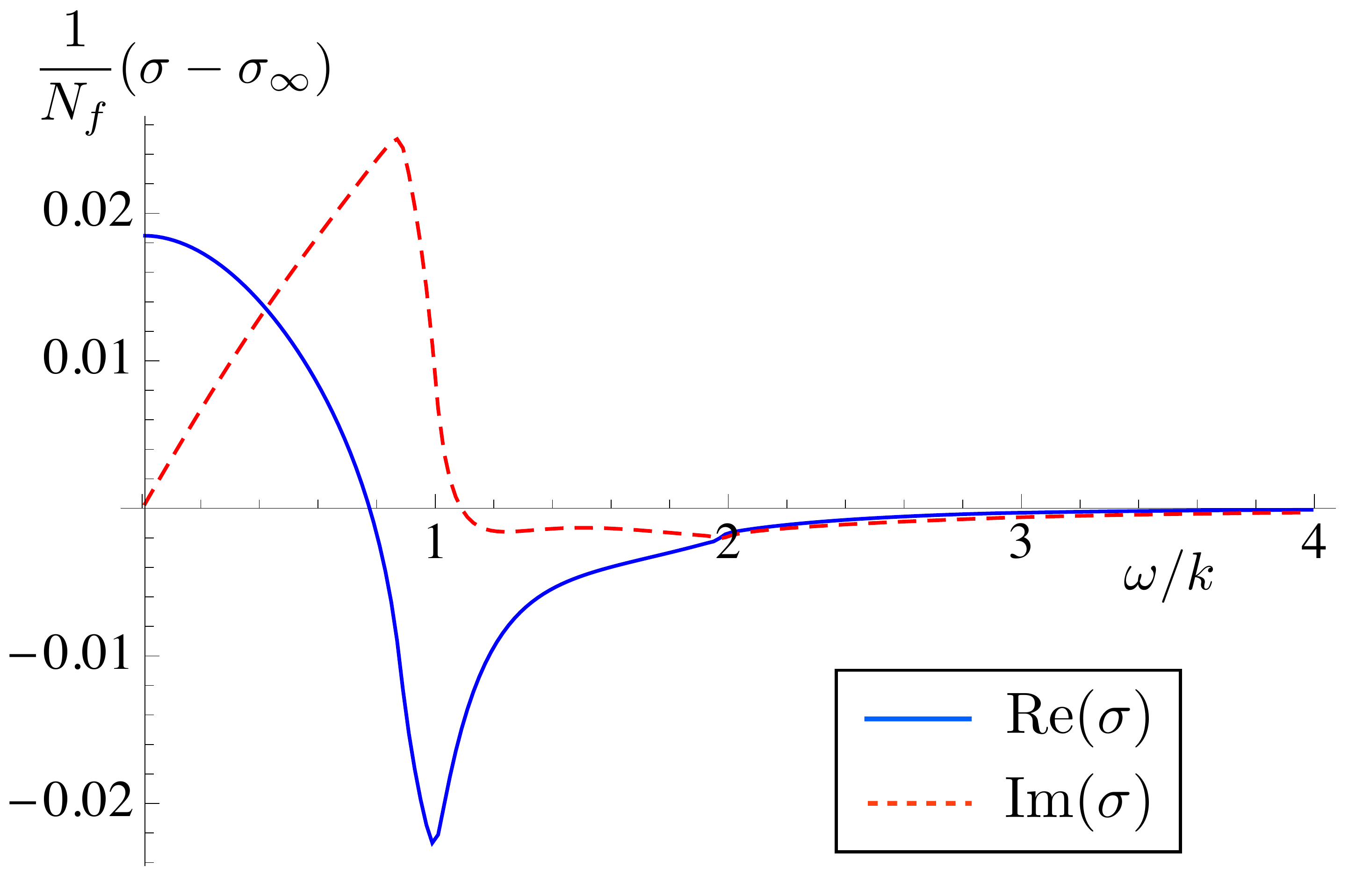}
\caption{Frequency-dependent conductivity at $V/k = 0.5$ for $N_f$ Dirac fermions in a periodic chemical potential.
Here $\sigma_\infty = \sigma (\omega \rightarrow\infty)$ and the Dirac CFT has $\sigma_\infty = N_f/16$.}
\label{fig:sigma0p5}
\end{center}
\end{figure}

\begin{figure}[H]
   \centering
   \includegraphics[width=4.5in]{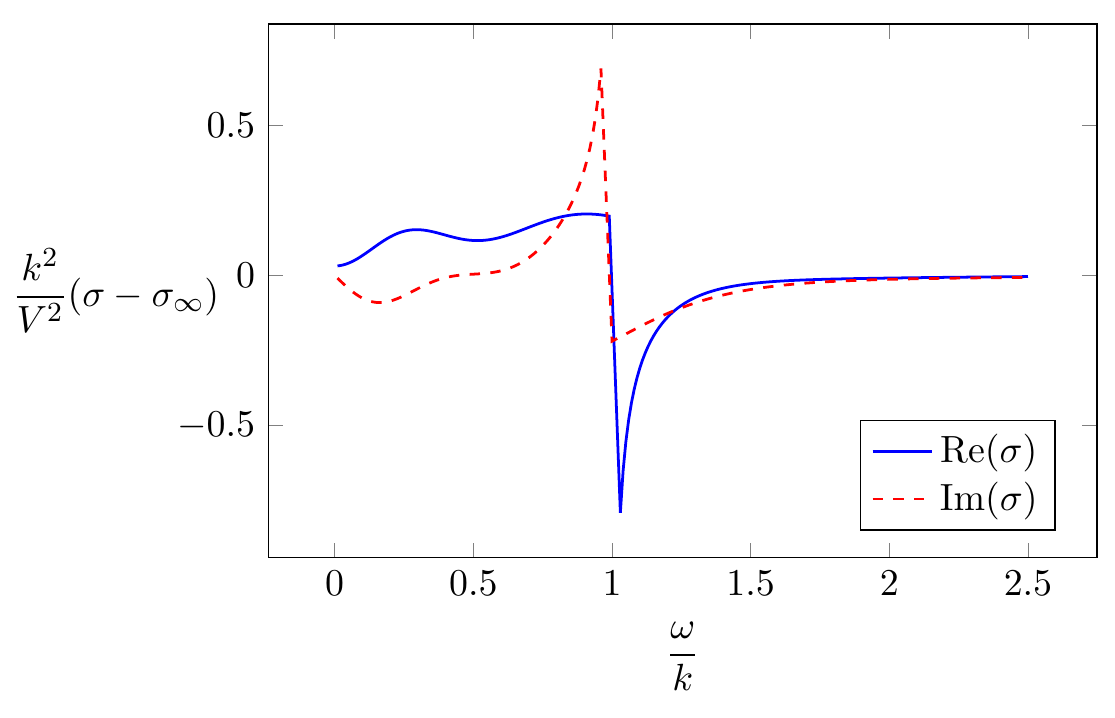}
   \caption{A plot of the analytic solution from holography 
   for $\mathrm{Re}(\sigma)$ vs. $\omega$ in the perturbative regime $V \ll k$, normalized to emphasize the strength of the perturbations.}
   \label{fig:sigmahol1}
\end{figure}

Turning on a potential, 
we show the result at small $V/k$ for free Dirac fermions in Fig.~\ref{fig:sigma0p5}, and that from holography in Fig.~\ref{fig:sigmahol1}. 
Note the remarkable similarity in the basic features of the frequency dependence.
The correspondence for larger $V/k$ is not as complete, but the two methods do share the common feature of having a peak in $\mbox{Re}[\sigma (\omega)]$
at $\omega \sim k$, followed by a dip until $\omega \sim V$, as we will see in Sections~\ref{sec:holo} and~\ref{sec:graphene}.
A `resonance' at $\omega \sim k$ also appeared in the $\mu_0 \neq 0$ results of Ref.~\cite{gary2}.

\begin{figure}[H]
\begin{center}
\includegraphics[width=4in]{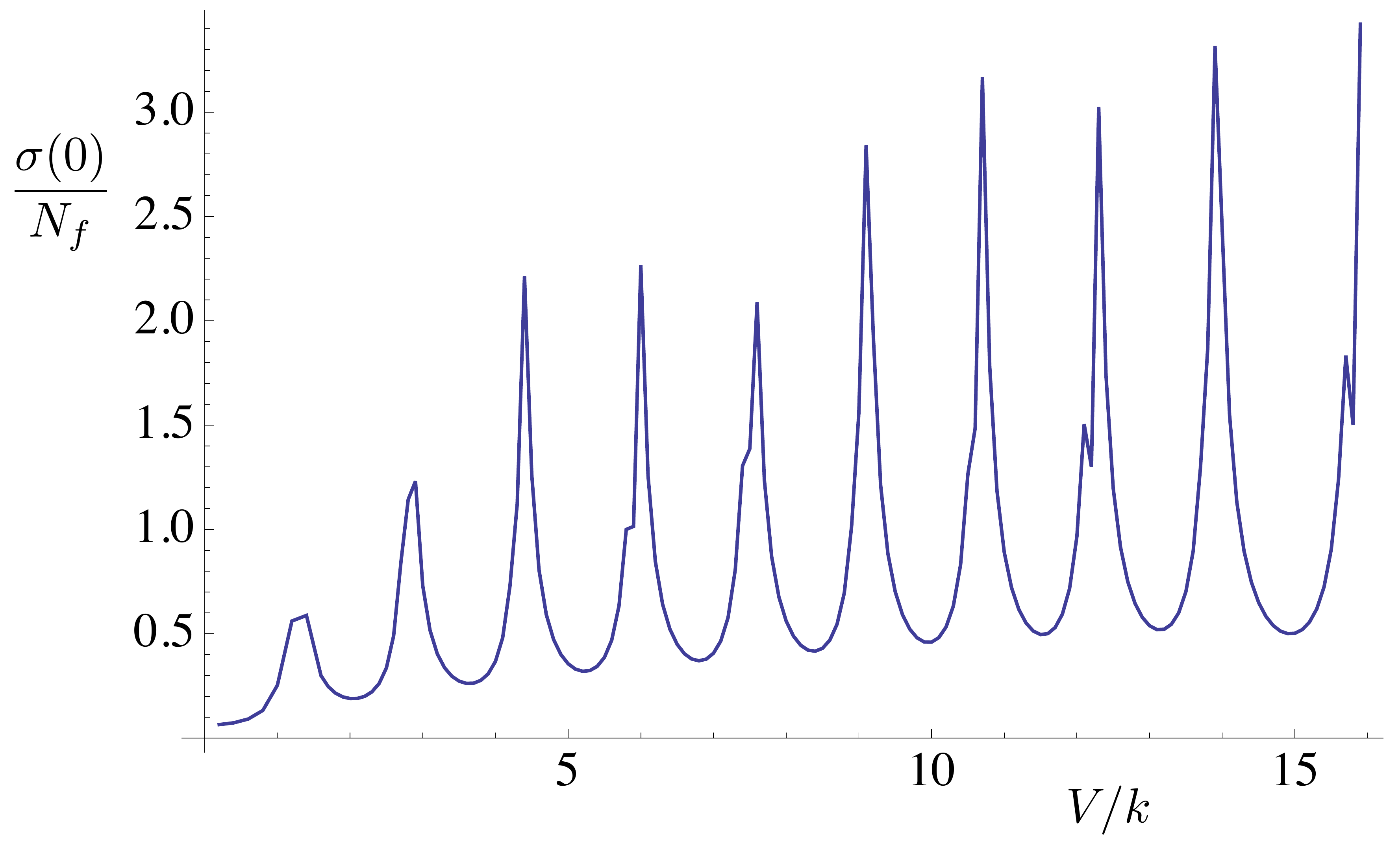}
\caption{D.C. conductivity for $N_f$ Dirac fermions in a periodic chemical potential as a function of $V/k$.}
\label{fig:sigmaV}
\end{center}
\end{figure}

\begin{figure}[H] 
   \centering
   \includegraphics[width=4in]{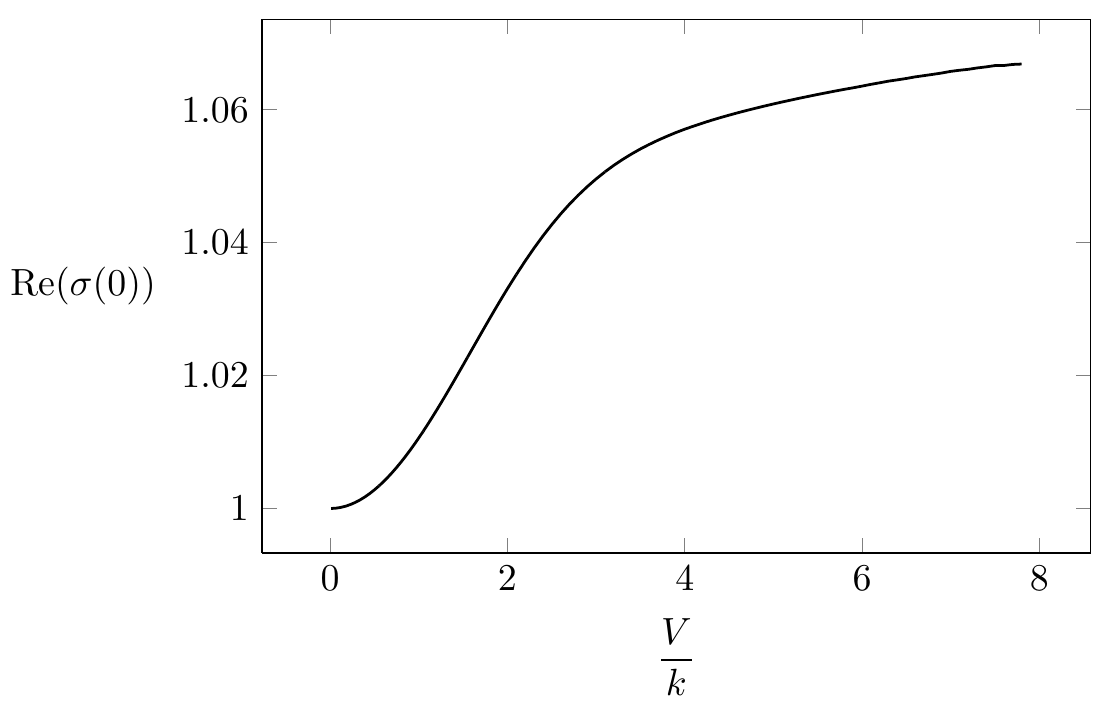}
   \caption{We show the holographic computation of $\sigma(0)$ (approximated by $\sigma(0.01)$, as our numerics cannot compute the d.c. conductivity directly) as a function of $V/k$.  This data was taken at $k/T=8$.}
   \label{fig:sigmaholV}
\end{figure}

Another interesting comparison between the two theories is in the $V/k$ dependence of the d.c. conductivity, $\sigma(0)$.
The result of the free Dirac fermion computation is shown in Fig.~\ref{fig:sigmaV} (a similar plot has appeared earlier in the graphene literature \cite{brey1}), while the result of holography is in Fig.~\ref{fig:sigmaholV}.
There are sharp peaks in $\sigma (0)$ in the Dirac fermion computation 
are at precisely the points where the universality class of the IR CFT changes {\em i.e.\/} at the values of $V/k$ where $N_D$ jumps by 2.
These transition points are described by a non-relativistic theory, where the Dirac fermion dispersion has the form in Eq.~(\ref{qcubed}).
It is evident that the holographic theory does not include the physics of the transition points and the local Fermi surfaces of the Dirac theory, 
and its IR theory evolves smoothly as
a function of $V/k$. However, if we smooth out the peaks in the Dirac fermion computation, we see that their average resembles
the evolution in the holographic theory as a function of $V/k$.

\section{Holography}
\label{sec:holo}

We begin by describing the simplest possible holographic description of a theory with a conserved U(1) charge placed in a periodic potential:   classical Einstein-Maxwell theory with a U(1) gauge field.    The metric is subject to the boundary condition that it is asymptotically $\mathrm{AdS}_4$ in the UV: \begin{equation}
\mathrm{d}s^2 = g^0_{\mu\nu} \mathrm{d}x^\mu \mathrm{d}x^\nu = \frac{L^2}{z^2} \left[\mathrm{d}z^2 - \mathrm{d}t^2 + \mathrm{d}x^2 + \mathrm{d}y^2\right], \text{  as } z\rightarrow 0.    \label{g0}
\end{equation}For much of the discussion, we will choose to rescale to $L=1$.    The gauge field is subject to the UV boundary condition \begin{equation}
\label{eq:A0bc}
\lim_{z \to 0} A_\mu \mathrm{d}x^\mu = V\cos(kx)\mathrm{d}t.
\end{equation}It is clear that both $V$ and $k$ are the only dimensional quantities in the problem which are relevant, and thus the dynamics of the theory can only depend on the ratio $V/k$.   

The equations of motion of the Lagrangian in Eq.~(\ref{EM}) are Einstein's equation \begin{equation}
R^{\mu\nu} - \frac{1}{2}Rg^{\mu\nu} + \Lambda g^{\mu\nu} = \frac{\kappa^2}{e^2}T^{(\mathrm{EM})\mu\nu}\label{eq:einstein},
\end{equation} 
where $T^{(\mathrm{EM})}_{\mu\nu}$ is the stress tensor of the U(1) field, \begin{equation}
T^{(\mathrm{EM})}_{\mu\nu} = {F_\mu}^\rho F_{\nu\rho} - \frac{1}{4}g_{\mu\nu} F^{\rho\sigma}F_{\rho\sigma},
\end{equation}and Maxwell's equations
 \begin{equation}
\nabla_\mu F^{\mu\nu} = 0.   \label{maxweq}
\end{equation}
Note that the U(1) stress tensor is traceless, in addition to having no divergence.   We will typically set $\kappa^2=1/2$ and $e=1$ as well for this paper.   Note that technically $\gamma = 2e^2L^2/\kappa^2$ \cite{hartnollcqg} forms a dimensionless quantity, but as this is usually O(1), we will content ourselves with the choice $\gamma=4$, and neglect to include this factor in our calculations.

Let us briefly outline the remainder of this section.   We will begin by briefly describing our numerical methods, which require us to extract zero temperature numerical results from finite temperature.    Then we will describe the dynamics when $V\ll k$:  this corresponds to a limit where the theory is described by small perturbations around Einstein-Maxwell theory in a pure $\mathrm{AdS}_4$ background.     The final subsection describes the results when $V \gg k$, where we use simple scaling arguments and numerics to understand the results.

\subsection{Numerical Methods}

\newcommand{\z}{z}

For our numerical analysis  we employ a characteristic formulation of Einstein's equations.
Diffeomorphism and translation invariance
in the transverse direction allows the bulk metric to be written
\begin{equation}
\mathrm{d}s^2 = -A \mathrm{d}t^2
+ 2F \mathrm{d}x \mathrm{d}t  + \Sigma^2 \left(\mathrm{e}^B \mathrm{d}x^2 + \mathrm{e}^{-B}\mathrm{d}y^2\right)  - \frac{2\mathrm{d} \z \mathrm{d} t}{\z^2} .   \label{numericform}
\end{equation}
where the functions $A$, $B$, $\Sigma$ and $F$ all depend on the AdS radial coordinate $\z$, the spatial coordinate $x$ and time $t$.   The spatial coordinate $x$ is taken to be periodic with period $2\pi /k$.  
In this coordinate system lines of constant time $t$ constitute radial infalling null geodesics affinely parameterized by $1/\z$.  
At the AdS boundary, located at $\z = 0$, $t$ corresponds to time in the dual quantum field theory.  The metric (\ref{numericform}) is invariant under the residual 
diffeomorphism 
\begin{equation}
\label{eq:diffeo}
\frac{1}{\z} \to \frac{1}{\z} + \xi(t,x),
\end{equation}
where $\xi(t,x)$ is an arbitrary function.

For the gauge field we choose the axial gauge in
which $A_\z = 0$.  The non-vanishing components of the gauge field are then given by
\begin{equation}
A_\mu \mathrm{d}x^\mu = A_t \mathrm{d}t + A_x \mathrm{d}x.
\end{equation}
We therefore must solve the Einstein-Maxwell system (\ref{eq:einstein}) and (\ref{maxweq}) for the six functions $A$, $B$, $\Sigma$, $F$ $A_t$ and $A_x$.

At first sight the Einstein-Maxwell system (\ref{eq:einstein}) and (\ref{maxweq}) appears over determined: there are ten equations of motion
for six functions $A$, $B$, $\Sigma$, $F$, $A_t$ and $A_x$.  However, four of these equations are radial constraint equations and in a sense redundant.
The radial constraint equations are simply the radial components of both Einstein's equations (\ref{eq:einstein}) and Maxwell's
Equations  (\ref{maxweq}) (with all indices raised) and are first order in radial derivatives.  
Moreover, if the radial constraint equations are satisfied at one value of $\z$ then the remaining equations imply they will be satisfied at all $\z$.
Because of this, it is sufficient to impose the radial constraint equations as boundary conditions on the remaining equations of motion in the Einstein-Maxwell system.
We therefore take our equations of motion
to be the $(\mu, \nu) = (t,t)$, $(x,x)$, $(y,y)$ and $(0,x)$ components of Einstein's equations (\ref{eq:einstein}) and the $\mu = t$ and $\mu = \ x$ components 
of Maxwell's equations  (\ref{maxweq}), again with all indices raised.

Upon suitably fixing boundary conditions our characteristic formulation of the Einstein-Maxwell system yields a well-behaved system of 
partial differential equations with one important caveat.  As is easily verified Einstein's equations become singular when $\Sigma = 0$.
This happens when the congruence on infalling geodesics develop caustics. Indeed, $\Sigma^2$ is the local area
element of radial infalling light sheets.  Nevertheless this problem can be ameliorated by introducing a small background 
temperature $T$.   By introducing a small temperature
the geometry will contain an black hole with finite area.  By increasing $T$ the location of the black hole's event horizon can be pushed closer to the 
boundary and thereby envelop any caustics that may exists.  As the interior of the horizon is causally disconnected from the near-boundary geometry,
one need only solve the Einstein-Maxwell system up to the location of the horizon.  

It is insightful to solve the Einstein-Maxwell system with a power 
series expansion at the AdS boundary $\z = 0$.  In doing so we impose the boundary conditions that the boundary geometry is that of 
flat Minkowski space and that the gauge field asymptotes to $\lim_{\z \to 0} A_\nu \mathrm{d}x^\nu = \mu \mathrm{d}t+ a_x \mathrm{d}x$ where
$\mu$ is given in Eq.~(\ref{chemx}) and $a_x$ is arbitrary.  Imposing these boundary conditions we find the following asymptotic forms
\begin{subequations}
\label{eq:asymtotics}
\begin{align}
A &= \frac{1}{\z^2} \left ( 1 + 2 \xi \z + (\xi^2- 2 \partial_t \xi) \z^2 + A^{(3)} \z^3 + O(\z^4) \right ),
&B &=  B^{(3)} \z^3 + O(\z^4), \\
F &= \partial_x \xi + F^{(1)} \z + O(\z^2), 
&\Sigma &= \frac{1}{\z} + \xi + O(\z^3), \\
A_t &= \mu +  A_t^{(1)} \z + O(\z^2),
& A_x &=a_x +   A_x^{(1)} \z + O(\z^2),
\end{align}
\end{subequations}
where $A^{(3)}$, $F^{(1)}$,  $A_t^{(1)}$, $A_x^{(1)}$ are undetermined and hence sensitive to the bulk geometry.  
However, the aforementioned radial constraint equations imply that these coefficients satisfy a system of constraint equations
which are most easily written in terms of the expectation value of the boundary stress tensor and the expectation value of the boundary
current.  In terms of these expansion coefficients the 
expectation value of the boundary stress and current operators read \cite{skenderis}
\begin{subequations}
\label{eq:holorenorm}
\begin{align}
\langle T^{tt} \rangle &= -\frac{2}{3} A^{(3)}, 
&\langle T^{tx} \rangle &= - F^{(1)}, 
&\langle T^{xx} \rangle &= - \frac{1}{3} A^{(3)} + B^{(3)}, 
&\langle T^{yy} \rangle &= - \frac{1}{3} A^{(3)} - B^{(3)}, 
\\
\langle J^0 \rangle &= - A_t^{(1)},
&\langle J^x \rangle &= A_x^{(1)}+ \partial_x \mu,
\end{align}
\end{subequations}
with all other components vanishing.  The radial constraint equations then require
\begin{align}
\label{eq:bndcons}
\partial_i \langle T^{ij} \rangle &= \frac{2 \kappa^2}{3 e^2} \langle J_{i} \rangle f^{i j}, 
&\partial_i \langle J^i \rangle &= 0,
\end{align}
where $f_{ij} \equiv \lim_{\z \to 0} \left ( \partial_i A_j - \partial_j A_i \right)$ is the boundary field strength
and all boundary indices are raised and lowered with the Minkowski space metric
$\eta_{ij} = {\rm diag} (-1,1,1)$.  
The expansion coefficient $\xi$ is related to the residual diffeomorphism invariance in Eq. (\ref{eq:diffeo}) and is therefore 
arbitrary.

To compute the conductivity we let $a_x = (E_x/\mathrm{i} \omega)\, \mathrm{e}^{-\mathrm{i} \omega t}$ where the boundary electric field $E_x \to 0$ 
and study the induced current.
In this case the fields reduce to a static piece plus an infinitesimal time dependent perturbation
\begin{subequations}
\begin{align}
A(t,x,\z) &= \tilde A(x,\z) + E_x \, \delta A(x,\z) \mathrm{e}^{-\mathrm{i} \omega t},
&B(t,x,\z) &=  \tilde B(x,\z) + E_x \, \delta B(x,\z) \mathrm{e}^{-\mathrm{i} \omega t},
\\
\Sigma(t,x,\z) &= \tilde \Sigma(x,\z) + E_x \, \delta \Sigma(x,\z) \mathrm{e}^{-\mathrm{i} \omega t},
&F(t,x,\z) &= \tilde F(x,\z) + E_x \, \delta F(x,\z) \mathrm{e}^{-\mathrm{i} \omega t},
\\
A_t(t,x,\z) &= \tilde A_t (x,\z)+ E_x \, \delta A_t (x,\z) \mathrm{e}^{-\mathrm{i} \omega t},
&A_x(t,x,\z) &= \tilde A_x(x,\z) + E_x \, \delta A_x(x,\z) \mathrm{e}^{-\mathrm{i} \omega t}.
\end{align}
\end{subequations}
We therefore have to solve two systems of equations.  We first solve the non-linear
but static Einstein-Maxwell system for the tilded fields with $a_x =0$.   The static fields induce 
no current.  Next, we linearize the Einstein-Maxwell system in $E_x$
and solve the resulting linear system for the time dependent perturbations.  
Upon doing so and extracting $\langle J_x \rangle$ via (\ref{eq:holorenorm}) we define the conductivity by
\begin{equation}
\sigma(\omega) \equiv \frac{ \mathrm{e}^{\mathrm{i} \omega t}}{E_x}  \frac{k}{2\pi} \int\limits_{-\pi/k}^{\pi/k} \mathrm{d}x \langle J_x(t,x) \rangle.   \label{boeq1}
\end{equation}

We discretize both the static and linear Einstein-Maxwell systems using pseudospectral methods.  We decompose the $x$ dependence 
of all functions in terms of plane waves and the radial dependence in terms of Chebyshev polynomials.
We solve the static Einstein-Maxwell system by employing Newton's method.

\subsubsection{The Static Einstein-Maxwell System}
\label{sec:static}

For the static Einstein-Maxwell system we choose to impose the $(\mu, \nu) = (x,\z), \ (\z,\z)$ radial constraint components of Einstein's equations (\ref{eq:einstein})
as boundary conditions at the location of the event horizon.  Likewise, we also choose to impose the $\mu = \z$ radial constraint component
of Maxwell's equations (\ref{maxweq}) as a boundary condition at the horizon.  Furthermore, we exploit the residual diffeomorphism invariance in Eq.~(\ref{eq:diffeo})
by fixing the location of the event horizon to be at $\z = 1$.  The $(\mu, \nu) = (x,\z), \ (\z,\z)$ components of Einstein's equations and the 
$\mu = \z$ component of Maxwell's equations are satisfied at the horizon provided
\begin{align}
\label{eq:bcz1}
\tilde A &= 0, 
&\partial_{\z} \tilde A &= 4 \pi T,
&\tilde F &= 0,
&\tilde A_t &= 0,
\end{align}
at $\z = 1$.  The second boundary condition in (\ref{eq:bcz1}) comes from Hawking's formula relating the local surface gravity 
of the black hole to the temperature $T$ of the black hole.
We choose to impose the remaining radial constraint equation --- the $(\mu, \nu) = (t,\z)$ components of Einstein's equations --- as a boundary condition
at $\z = 0$.  Specifically, the $(\mu, \nu) = (t,\z)$ components of Einstein's equations is satisfied at the boundary provided 
\begin{equation}
\label{eq:bc0z}
\lim_{\z \to 0} (\tilde A -\tilde \Sigma^2) = 0,
\end{equation}
which clearly the asymptotic expansions in Eq. (\ref{eq:asymtotics}) satisfy.  

For our numerical analysis we choose to make the following set of field redefinitions
\begin{align}
\label{eq:fieldredef}
a &\equiv \tilde A - \tilde \Sigma^2, & b &\equiv \frac{\tilde B}{\z^2}, & s & \equiv \tilde \Sigma - \frac{1}{\z}.
\end{align}
From the asymptotic expansions Eq. (\ref{eq:asymtotics}) and Eq.~(\ref{eq:bc0z}) we therefore 
impose the boundary conditions 
\begin{align}
a &= 0, 
&b  &= 0, 
&\partial_{\z} s &= 0, 
&\partial_{\z} \tilde F  &= 0, 
&\tilde A_t &= \mu,
&\tilde A_x  &= 0,
\end{align}
at $\z = 0$.
Likewise we translate the boundary conditions at $\z = 1$ in Eq.~(\ref{eq:bcz1})
into the variables $a$, $b$ and $s$.

All told we impose a total of 10 radial boundary conditions for a set of 6 second order partial differential equations.
Naively it might seem that two additional boundary conditions are required.  However, careful analysis of Einstein's equations
and Maxwell's equations show that two of the equations become first order at the horizon.  Therefore only 10 radial 
boundary conditions are required.

\subsubsection{Linearized Fluctuations}

Following Eqs.~(\ref{eq:fieldredef}) we choose to make the following field redefinitions for the time-dependent linear perturbations
\begin{align}
\label{eq:fieldredefline}
\delta a &\equiv \delta A - 2 \tilde \Sigma \delta \Sigma, & \delta b &\equiv \frac{\delta B}{\z^2}.
\end{align}
In contrast to the static problem, where most of the radial constraints where imposed as boundary conditions at the horizon,
for the linear perturbations we choose to impose all the radial constraint equations as boundary conditions
at the AdS boundary.  This is tantamount to demanding that the asymptotic expansions (\ref{eq:asymtotics})
are satisfied near $\z = 0$ and that the conservations equations (\ref{eq:bndcons}) are satisfied.  In the $E_x \to 0$ limit the asymptotic expansions (\ref{eq:asymtotics})
and the conservation equations (\ref{eq:bndcons}) yield the mixed boundary conditions at $\z = 0$,
\begin{subequations}
\label{eq:linbc}
\begin{align}
2 \mathrm{i} \omega \partial_\z \delta a - 3 \partial_x \partial_z \delta F - \frac{2 \kappa^2}{e^2} \partial_x \tilde A_t \partial_\z \delta A_x -   \frac{2 \kappa^2}{e^2}  \mathrm{i} \omega \left ( \partial_\z \tilde A_x + 2 \partial_x \tilde A_t \right ) \delta A_x &= 0,
\\
\partial_x \partial_\z \delta a - 3 \partial_x \partial_\z \delta b - 3 \mathrm{i} \omega \partial_\z \delta F -  \frac{2 \kappa^2}{e^2}  \partial_x \tilde A_t \partial_z \delta A_t - \mathrm{i} \omega  \frac{2 \kappa^2}{e^2}  \partial_z \tilde A_t \delta A_x &= 0,
\\
\mathrm{i} \omega \partial_\z \delta A_t + \partial_x \partial_\z \delta A_x + \mathrm{i} \omega \partial_x \delta A_x &= 0,
\end{align} 
\begin{align}
\delta a & = 0, 
&\delta b &= 0,
& \delta \Sigma &= 0,
& \partial_\z \delta \Sigma &= 0,
& \delta f & = 0,
&\delta A_t &= 0,
&\delta A_x &= \frac{1}{\mathrm{i} \omega}.
\end{align}
\end{subequations}

As above in Section~\ref{sec:static}, all told we impose a total of 10 radial boundary conditions for a set of 6 second order partial differential equations.
Again, careful analysis of linearized Einstein-Maxwell system shows that two of the equations become first order at the horizon.  Therefore only then 10 radial 
boundary conditions in Eq. (\ref{eq:linbc}) are required.

\subsection{The Perturbative Regime:  $V\ll k$}
When $V\ll k$ the geometry only slightly deviates from AdS and the conductivity only slightly deviates from a constant.    We thus perform an analytic calculation of the back-reacted metric, and the longitudinal conductivity, to lowest nontrivial order, $(V/k)^2$.     Because the analytic calculations are straightforward but cumbersome, we have placed them in Appendix \ref{appa}.   The results are outlined here, along with a comparison to our numerics.
For our analytic calculation we employ coordinates where the unperturbed metric is given by Eq.~(\ref{g0}).

\subsubsection{Geometric Results}
 It is trivial to compute the perturbative solution to Maxwell's equations around an AdS background geometry.   In $\mathrm{AdS}_4$, choosing the covariant Lorentz gauge $\nabla^\mu A_\mu = 0$, Eq. (\ref{maxweq}) simplifies to\begin{equation}
 \nabla_\rho\nabla^\rho A_\mu = 0.
 \end{equation}
 For the AdS$_4$ geometry given by (\ref{g0}) the above wave equation further simplifies to
 \begin{equation}
 0= \left(\partial_z^2 - \partial_t^2 + \partial_x^2 +\partial_y^2 \right) A_t.
 \end{equation}
 The static solution satisfying the boundary condition (\ref{eq:A0bc}) is then
 \begin{equation}
A_\mu \mathrm{d} x^\mu  = V \cos(kx) \mathrm{e}^{-kz} \mathrm{d}t.  \label{bgf}
\end{equation} To compute the perturbation in the geometry induced by the above gauge field we decompose the metric as 
\begin{equation}
g_{\mu\nu} = g^0_{\mu\nu}  + h_{\mu\nu}
\end{equation}where $g^0_{\mu\nu}$ is the pure AdS metric given by Eq. (\ref{g0}), and $h_{\mu\nu}$ is the linearized metric response to lowest order.   
We compute $h_{\mu\nu}$ by solving the graviton's linearized equation of motion, sourced by the stress tensor of the background gauge field given by Eq. (\ref{bgf}).    This is simple to find using a transverse traceless gauge, and we simply cite the solution here:\begin{subequations}\begin{align}
h_{zz}(z,x) &= H_{zz}(z)\cos(2kx), \\
h_{zx}(z,x) &= H_{zx}(z)\sin(2kx), \\
h_{xx}(z,x) &= H_{xx}(z)\cos(2kx), \\
h_{yy}(z,x) &=  G_{yy}(z) + H_{yy}(z)\cos(2kx), \\
h_{tt}(z,x) &= G_{yy}(z)-H_{yy}(z)\cos(2kx),\\
h_{tx} &= h_{ty} = h_{xy} = h_{zt} = h_{zy} = 0.
\end{align}\end{subequations}We know exactly:\begin{subequations}\begin{align}
G_{yy}(z) &= V^2 \frac{1-\mathrm{e}^{-2kz}\left(1+2kz+2k^2z^2\right)}{16k^2z^2}, \\
H_{zz}(z) &= \frac{V^2}{16}\left[\mathrm{e}^{-2kz}\left(2+(1+2kz)\log(kz)\right) + (2kz-1)\mathrm{e}^{2kz}\mathrm{Ei}(-4kz)\right] \notag \\
&\;\;\;\; + \frac{V^2\left(\log 4 + \gamma - 2\right)}{16}(1+2kz) \mathrm{e}^{-2kz}.
\end{align}\end{subequations} and where \begin{subequations}\begin{align}
H_{zx} &= -\frac{1}{2k}\left(\frac{\mathrm{d}}{\mathrm{d}z} - \frac{2}{z}\right)H_{zz}(z), \\
H_{xx} &= \frac{1}{2k}\left(\frac{\mathrm{d}}{\mathrm{d}z} - \frac{2}{z}\right)H_{zx}(z), \\
\frac{\mathrm{d}^2 H_{yy}}{\mathrm{d}z^2} + \frac{2}{z} \frac{\mathrm{d}H_{yy}}{\mathrm{d}z} - \left(\frac{2}{z^2} + 4k^2\right) H_{yy} &= - \frac{2H_{zz}}{z^2}
\end{align}\end{subequations} The boundary conditions on $H_{yy}$ are that it vanish at $z=0$ and $z=\infty$, as do all perturbations.   The UV asymptotic behavior of this metric is simply that of AdS, with all corrections to the UV metric of order $r$, while in the IR, it is a rescaled AdS: \begin{equation}
\mathrm{d}s^2 (z\rightarrow \infty) = \frac{1}{z^2}\left[\mathrm{d}z^2 + \mathrm{d}x^2 + \left(1+\left(\frac{V}{4k}\right)^2\right) \mathrm{d}y^2 - \left(1-\left(\frac{V}{4k}\right)^2\right) \mathrm{d}t^2  \right].
\end{equation}The rescaling only affects the $t$ and $y$ directions -- the overall radius $L$ of the AdS space has not renormalized.   Compared to the UV, we thus have the effective length scale in the perpendicular direction to the spatial modulation slightly larger: \begin{equation}
\frac{L_y^{\mathrm{IR}}}{L_y^{\mathrm{UV}}} = \sqrt{\frac{g_{yy}(z=\infty)}{g_{yy}(z=0)}} \approx 1 + \frac{1}{2}\left(\frac{V}{4k}\right)^2.   \label{relativeLy}
\end{equation}Right away, we suspect that the resulting IR AdS geometry implies that the IR effective theory is also conformal. We will see evidence for this when we compute the conductivity.   

\begin{figure}[here] 
   \centering
   \includegraphics{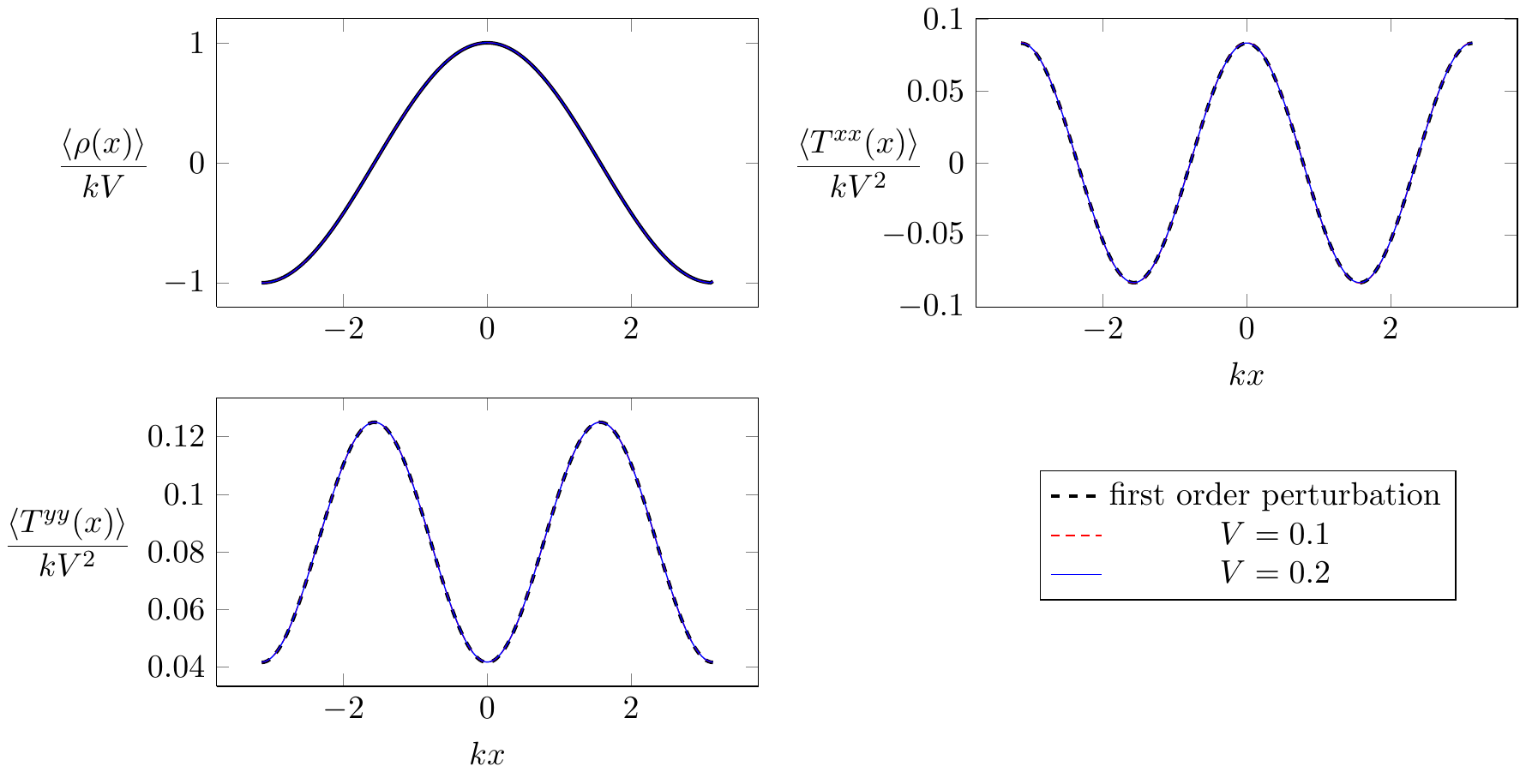}
   \caption{We compare the expectation values of simple operators in the CFT between our first-order perturbative expressions and numerical results.   In the numerical results, we have subtracted off the thermal contribution to the stress tensor given by Eq. (\ref{thermalstress}).  As is clear, we find excellent agreement with the numerics.}
   \label{rhostress}
\end{figure}

From the asymptotic of the gauge field and metric perturbation we extract the expectation value of the charge density and the stress tensor via Eq.~(\ref{eq:holorenorm}).
This requires a trivial change of coordinates near the boundary to put the AdS$_4$ metric (\ref{g0}) into the form (\ref{numericform}).
From the asymptotics of the gauge field we obtain the expectation value of the charge density $\rho$
\begin{equation}
\langle \rho(x)\rangle = Vk\cos(kx).
\end{equation}
From the asymptotic of the metric perturbation we obtain 
\begin{subequations}\label{expstress}\begin{align}
\langle T_{xx}(x)\rangle &= \frac{V^2 k}{12}\cos(2kx), \\
\langle T_{yy}(x)\rangle &= \frac{V^2 k}{12} - \frac{V^2k}{24}\cos(2kx), \\
\langle T_{tt}(x)\rangle &= \langle T_{xx}(x)+T_{yy}(x)\rangle, \\
\langle T_{tx}(x) \rangle &= \langle T_{ty}(x)\rangle = \langle T_{xy}(x)\rangle = 0.
\end{align}\end{subequations}
In Figure~\ref{rhostress} we compare the above analytic formulas to our numerics at small $V$ and find excellent agreement
when thermal contributions to the numerical stress tensor are subtracted off.  This is reasonable because 
in the $T \to 0$ and $V \to 0$ limit the thermal and potential contributions to $\langle T_{ij}\rangle$ do not mix.
The thermal stress tensor is simply  
\begin{equation}
2\langle T^{\text{thermal}}_{xx}\rangle = 2\langle T^{\text{thermal}}_{yy}\rangle = \langle T_{tt}^{\text{thermal}} \rangle = \frac{2}{3} \left(\frac{4\pi T}{3}\right)^3. \label{thermalstress}
\end{equation}

\subsubsection{Conductivity}

\begin{figure}[here] 
   \centering
      \includegraphics{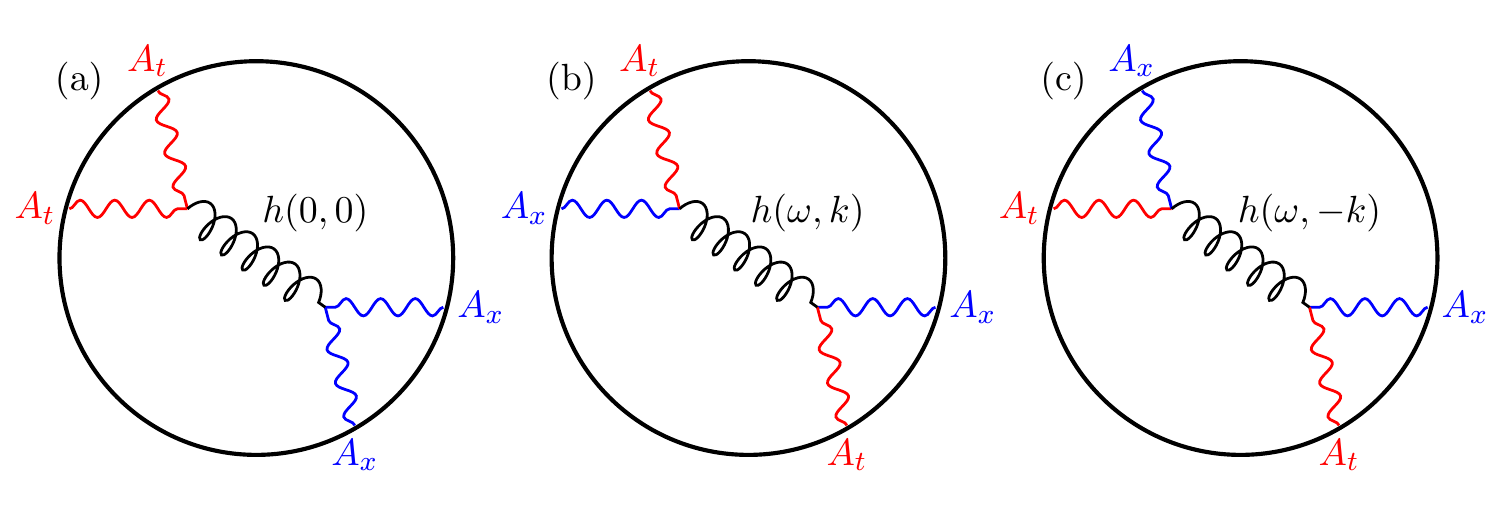} 
   \caption{The three Witten diagrams required to compute the conductivity.  (a) shows the elastic channel, and (b-c) show the inelastic channel.   The internal momenta in the $t,x$ direction carried by the graviton is also shown.}
   \label{witten}
\end{figure}

To compute the conductivity $\sigma$ we compute three Witten diagrams, as shown in Figure \ref{witten}.   In each Witten diagram, there are two probe photons $A_x$, which are used to measure the conductivity, which scatter at tree level off of two background photons $A_t$ via a single graviton $h$.    As is shown in the figure, these diagrams break up into an elastic channel where the probe gauge field scatters off of modified geometry, and an inelastic channel where the probe gauge field scatters off of the background gauge field.    After a cumbersome but straightforward calculation, one can find that \begin{equation}
\sigma(\omega) =  \sigma_{\mathrm{el}}(\omega)  + \sigma_{\mathrm{inel}}(\omega)
\end{equation}where \begin{subequations}\label{2sigma}\begin{align}
\sigma_{\mathrm{el}}(\omega) &= \frac{V^2 k}{32(k-\mathrm{i}\omega)^3}, \\ 
\sigma_{\mathrm{inel}}(\omega) &= \mathrm{i}V^2k^{10}\frac{(k-\mathrm{i}\omega) P_1 (-\mathrm{i}\omega/k) - \sqrt{k^2-\omega^2-\mathrm{i}\epsilon} P_2(-\mathrm{i}\omega/k)}{32\omega^3 \left(k^2-\omega^2-\mathrm{i}\epsilon\right) (k-\mathrm{i}\omega)^8}
\end{align}\end{subequations}where $\epsilon$ is an infinitesimal quantity,  the principal square root is taken, and $P_1(x)$ and $P_2(x)$ are large polynomials defined as:\begin{subequations}\begin{align}
P_1(x) &= 2+14x+49x^2+119x^3+216x^4 +313x^5+338x^6 \notag \\
&+313x^7+216x^8+119x^{9}+49x^{10}+14x^{11}+2x^{12},\\
P_2(x) &= 2+16x+62x^2+160x^3+310x^4 +464x^5+532x^6 \notag \\
&+464x^7+310x^8+160x^{9}+62x^{10}+16x^{11}+2x^{12} .
\end{align}\end{subequations} We note that both the elastic and inelastic channels separately obey sum rules \cite{sachdevww}: \begin{equation}
0 =  \int\limits_0^\infty \mathrm{d}\omega \mathrm{Re}(\sigma_{\mathrm{el}})  =  \int\limits_0^\infty \mathrm{d}\omega \mathrm{Re}(\sigma_{\mathrm{inel}}).
\end{equation}

Although it may not be obvious from Eq. (\ref{2sigma}), it is clear physically (and true in perturbation theory) that for $\omega\gg k$, we have $\sigma(\omega)=1$, which is the pure AdS result.    Sometimes we will also refer to this value as $\sigma_{\infty}$, when we wish to emphasize the relative scaling of the conductivity in various frequency regimes.   At large energy scales, the geometry has not deformed substantially and therefore the perturbation lives in a pure AdS spacetime, which explains the UV behavior.    In the deep IR, the spacetime again returns to AdS, but compared to the UV an anisotropy has formed.    We find that the d.c. conductivity becomes \begin{equation}
\sigma(0) = 1+\frac{1}{2}\left(\frac{V}{4k}\right)^2.
\end{equation}This can be explained given our previous results for the geometry.   In the IR, the relative anisotropy in the metric can be rescaled away into $t$ and $y$ to give a pure $\mathrm{AdS}_4$ geometry, but at the consequence of rescaling the length scales $L_x$ and $L_y$ associated to the theory.   This means that relative to the UV CFT, the IR CFT conductivity will pick up a relative rescaling factor of \begin{equation}a
\frac{\sigma(0)}{\sigma_\infty} = \frac{L_y^{\mathrm{IR}}}{L_x^{\mathrm{IR}}} = 1+\frac{1}{2}\left(\frac{V}{4k}\right)^2,
\end{equation}using Eq. (\ref{relativeLy}).    Also, since the conductivity has not picked up any overall extra scaling factors, the charge $e$ associated to the U(1) gauge field has not renormalized (at least to this order in $V/k$).

A second interesting feature of the analytic result is a singularity at $\omega = k$.   For $\omega$ very close to $k$, we have \begin{equation}
\sigma(\omega) \sim \frac{\mathrm{i}}{\sqrt{k- \omega - \mathrm{i}\epsilon}} + \text{subleading terms}.
\end{equation}Note that for $\omega$ just larger than $k$, the perturbative expression implies $\mathrm{Re}(\sigma) < 0$, suggesting that \emph{perturbation theory has broken down}.    Physically, this singularity is due to on-shell gravitons in scattering events where the probe photon scatters off a background photon.   We note that the Euclidean time solution (see Eq. (\ref{eucsigmainel})) is perfectly well-behaved, and this singularity comes from analytically continuing a function with branch cuts.   

\begin{figure}[H]
   \centering
   \includegraphics[width=6in]{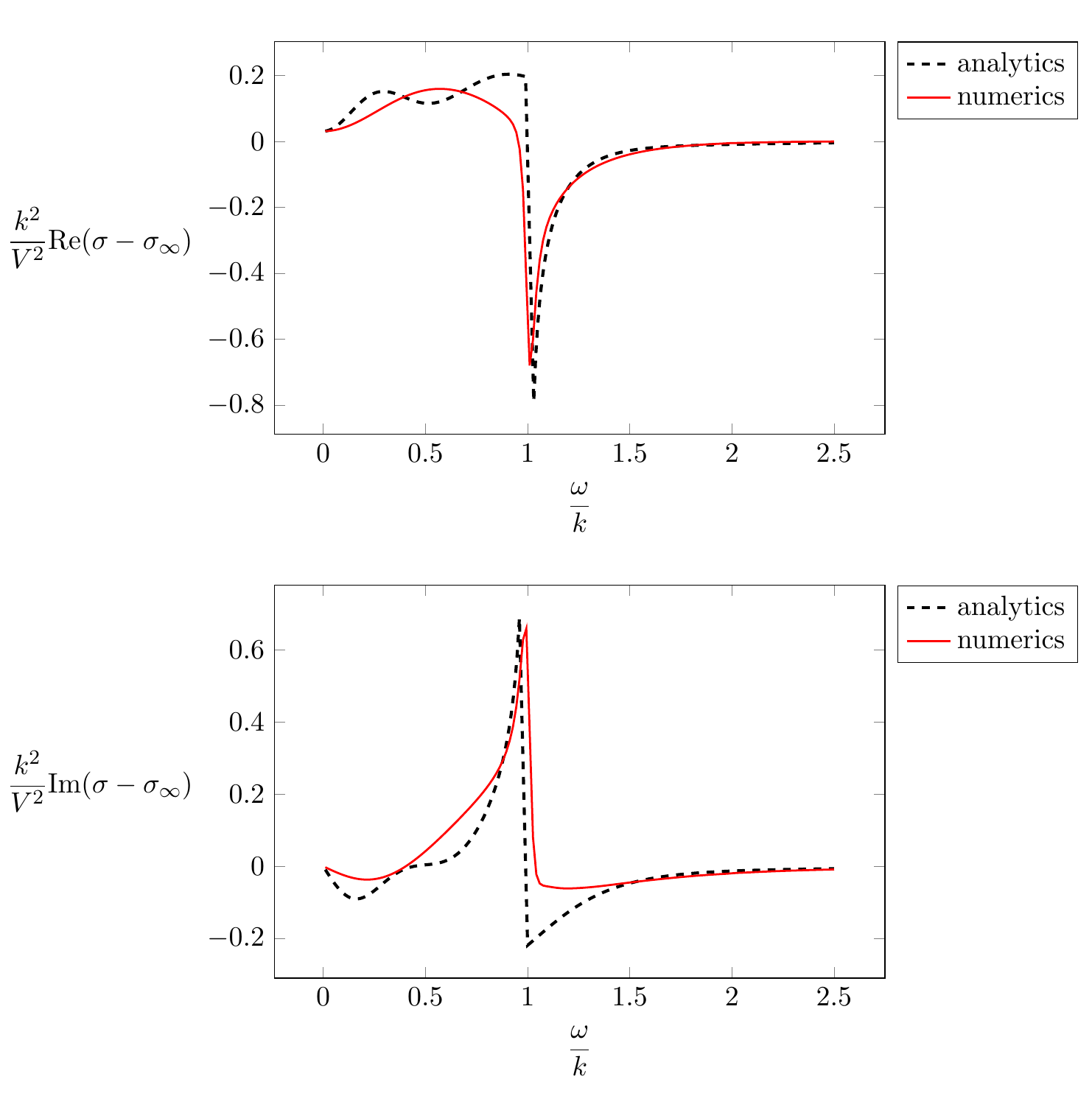}
   \caption{A plot of the analytic and numerical solution for $\mathrm{Re}(\sigma)$ vs. $\omega$ in the perturbative regime $V \ll k$.   For the numerical solution we took $k=1$, $T=1/50$, and $V=1/4$.  At these values the analytical and numerical values for $\langle T_{ij}\rangle$ agree to 3 orders of magnitude, placing us well in the perturbative regime.  The discrepancy at low frequencies is a consequence of the fact the series expansion of $\sigma$ in $T$ is not expected to converge near $\omega=k$ due to a non-analytic singularity;  as such, we expect this discrepancy to propagate an O($k$) distance in $\omega$. }
   \label{ancomp}
\end{figure}

The presence of this singularity makes a numerical evaluation of the conductivity quite tricky at finite temperature, which smooths out the singularity.  We show our results in Figure \ref{ancomp}.    Although there are large quantitative differences for $\omega<k$ between the analytical expression and the numerical result, we do note that the qualitative features, \emph{e.g.}  singular behavior at $\omega=k$ and the d.c. conductivity, do agree well, so we conclude that the finite temperature numerics are perfectly adequate to determine qualitative features of the conductivity.

\subsection{The Nonperturbative Regime: $V\gg k$}
Let us now discuss the nonperturbative regime:  $V\gg k$.    This is substantially more challenging to understand analytically, and indeed we will not be able to find an analytic solution.   Although the standard approach thus far has been to simply use numerical methods, we will see that simple heuristics also give us a surprising amount of insight into the nature of the nonperturbative geometry, and its consequences on the conductivity.    
\subsubsection{Geometric Results}
As with the perturbative regime, let us begin by describing the geometry when $V\gg k$.    As the geometry encodes many features of the physics, making sure that we have a good description of the nonperturbative geometry is crucial to understanding the conductivity.   In particular, we will want to understand the \emph{scales} at which the effects of the periodic potential become nonperturbative in the geometry.     The typical approach which has been attempted in these sorts of nonperturbative AdS problems is a matched asymptotic expansion: see \emph{e.g.} \cite{basu, dias}.   However, such approaches are not applicable here because we do not formally know the appropriate boundary conditions to impose in the IR.    Furthermore, even with a reasonable guess, the intermediate scale at which the spatial modulation significantly warps the geometry is mysterious.   We will have to resort to a more heuristic argument.

Let us consider the limit when $k\rightarrow 0$, and consider the geometry near $x=0$.   On a length scale which is small compared to $1/k$, the geometry will look like an extremal $\mathrm{AdS}_4$-Reissner-N\"ordstrom ($\mathrm{AdS}_4$-RN) black hole describing a CFT at chemical potential $V$.    We know that the metric undergoes nonperturbative corrections when placed in a background with a constant $A_t$ on the boundary \cite{hartnollcqg, chamblin}.   Let us make the \emph{ansatz} that in the UV, the spatial modulation is a minor correction, leading to a separation of length scales.   Such a separation of scales immediately allows us to write down a guess for the metric: \begin{equation}
\mathrm{d}s^2_{\mathrm{BO}} = \frac{1}{z^2} \left[-2\mathrm{d}z\mathrm{d}t - f(z,x)\mathrm{d}t^2 + \mathrm{d}x^2 + \mathrm{d}y^2 \right] \label{ds2bo}
\end{equation}where \begin{equation}
f(z) \equiv 1-4\left(\frac{z}{z_+(x)}\right)^3 + 3\left(\frac{z}{z_+(x)}\right)^4,
\end{equation}and \begin{equation}
z_+(x) \equiv \frac{\sqrt{12}}{V|\cos(kx)|}.
\end{equation}We will refer to Eq. (\ref{ds2bo})  as the Born-Oppenheimer metric 
(note that this Born-Oppenheimer ansatz can also be extended to the $\mu_0 \neq 0$ case).
We have used Eddington-Finkelstein coordinates here, for reasons which will become clear a bit later.    Correspondingly, the gauge field should be 
\begin{equation}
A_\mu \mathrm{d} x^\mu = V\cos(kx) \left[1-\frac{z}{z_+(x)}\right] \mathrm{d}t.
\end{equation}
These are exact non-perturbative solutions to Einstein-Maxwell theory so long as $x$-derivatives can be ignored.  An analytic treatment of this problem when long-wavelength deviations from a constant chemical potential are perturbative can be found in \cite{maeda}, but in our situation the deviations are nonperturbative as well, as we will see below. 
Just as locally boosted black brane geometries serve as a 
starting point for the fluid/gravity gradient expansions in \cite{bmw}, Eq. (\ref{ds2bo}) can serve as 
as starting point for a gradient expansion solution the Einstein-Maxwell system when $\partial_x \mu \to 0$, or equivalently in our case when $k \to 0$.
Just as in the case of the fluid/gravity gradient expansion, a requirement for Eq. (\ref{ds2bo}) to be a good starting point for a gradient expansion solution to the Einstein-Maxwell 
system is that there exists a wide separation of scales and that the dimensionless 
expansion parameter
\begin{equation}
\label{eq:expansionparm}
k z_+(x) \to 0.
\end{equation} 
This condition is satisfied simply by taking $k/V \to 0$, but it fails in the vicinity of the `turning points' $x = (n+1/2){\pi}/{ k}$, where $n$ is
any integer.    
We will shortly return to this point.

\begin{figure}[h] 
   \centering
      \includegraphics{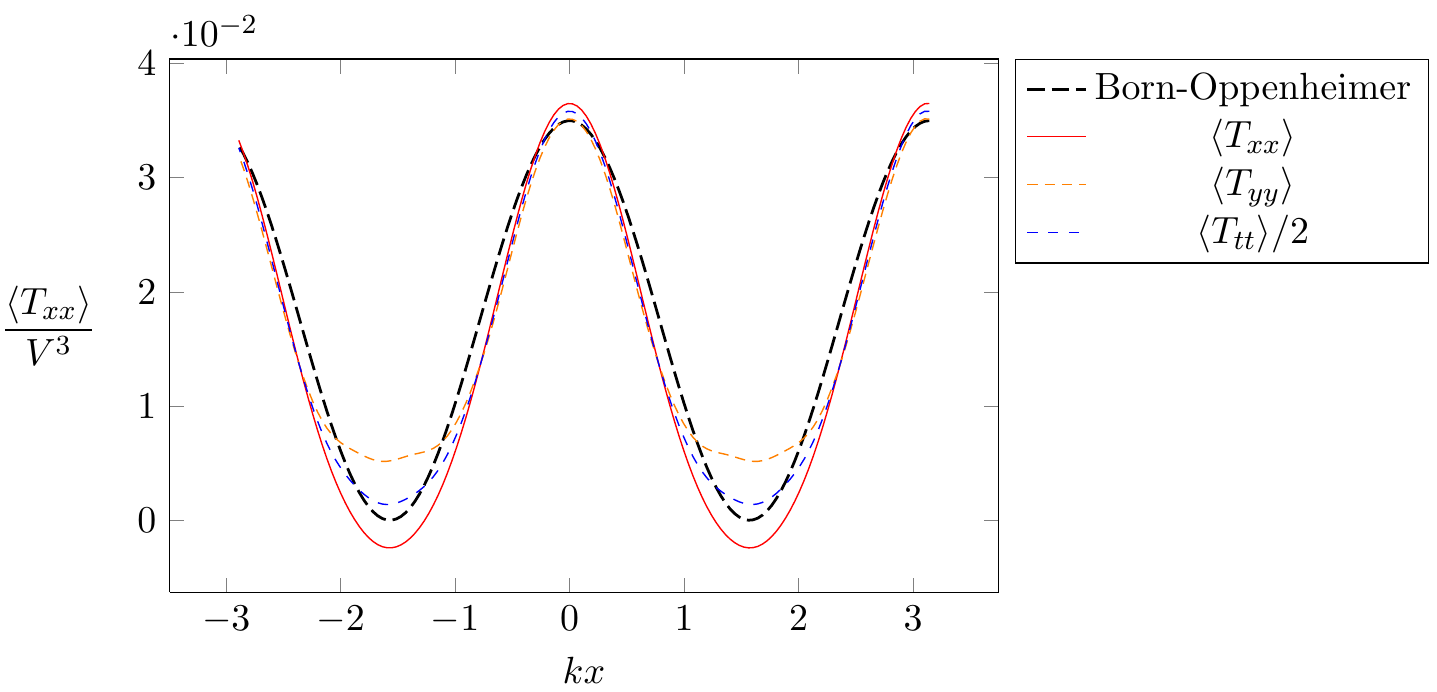}
         \caption{We show a comparison of the diagonal components of the stress tensor on the boundary to the results predicted by the Born-Oppenheimer approximation.   Away from the cross-over regions, agreement between all curves is within 10\%. Numerical data taken at $T=1/8$, $k=1$ and $V=10$.}
   \label{BOfig1}
 \end{figure}

The Born-Oppenheimer ansatz can be quantitatively checked in a gauge-invariant way by looking at the expectation value of the stress tensor on the boundary.   Looking back to Eq. (\ref{eq:holorenorm}), we see that the leading order $z^3$ term in $f(z,x)$ corresponds to the term which will contribute to the diagonal components of $\langle T_{ij}\rangle$.   Since the only scale in a RN black hole is $\mu$, we conclude that $\langle T_{ij}\rangle \sim \mu^3$:  a more precise check yields: \begin{equation}
\langle T_{tt}(x)\rangle = 2\langle T_{xx}(x)\rangle = 2\langle T_{yy}(x)\rangle = \left(1+\frac{z_+(x)^2\mu(x)^2}{4}\right)\frac{2}{3z_+(x)^3}
\end{equation}At zero temperature we have $z_+(x)\mu(x) = \sqrt{12}$, but for our numerics which occur at finite temperature the distinction matters.   We show the results of a comparison of these three stress tensors to the geometry when $V\gg k$ in Figure \ref{BOfig1}.   Despite our simple separation of length scales ansatz, this metric is \emph{quantitatively correct} in the UV.  Note that this approximation only fails near points where $\cos(kx)=0$.   For reasons which become clear in the next paragraph, this region is dominated by different physics, and we will refer to this as the ``cross-over region" in the UV geometry.

\begin{figure}[h] 
   \centering
      \includegraphics{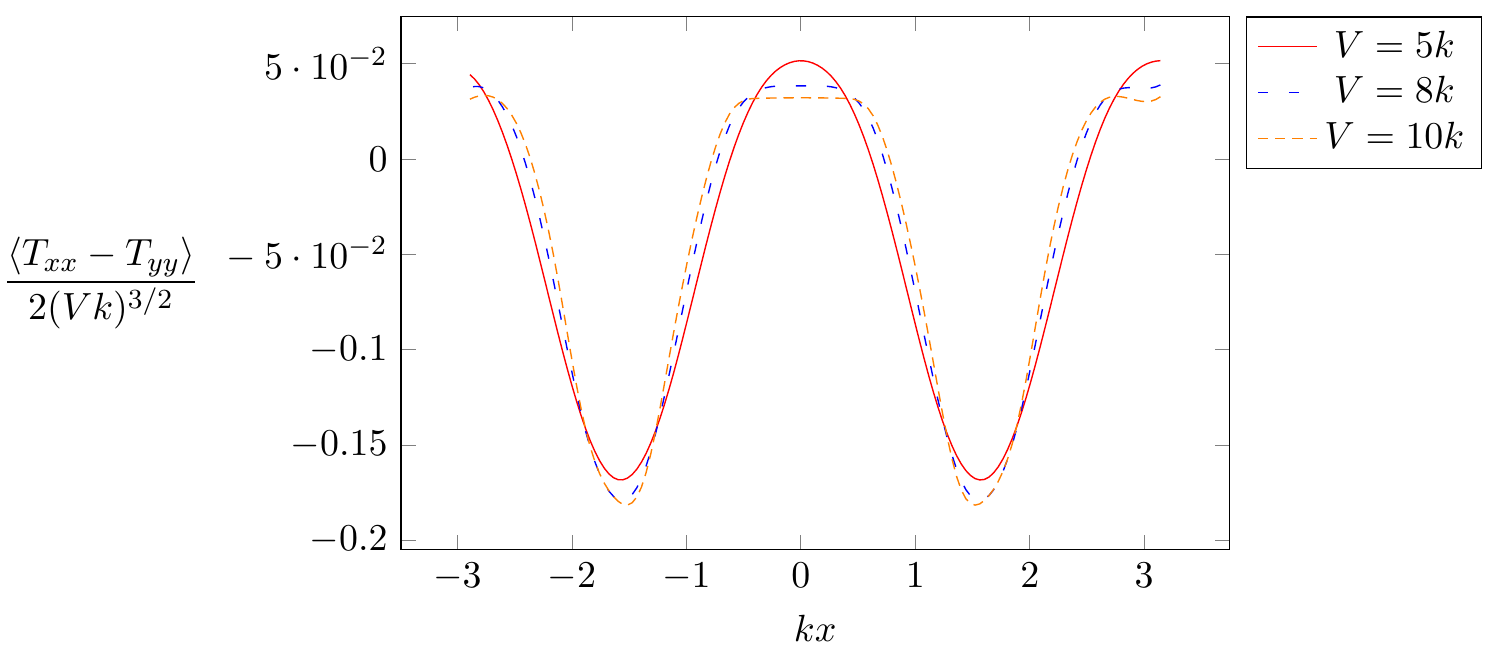}
         \caption{We show the anisotropy in the metric near the cross-over regions by looking at the stress tensor on the boundary.  Numerics were run at $T=1/8$, $k=1$ and varying $V$.}
   \label{BOfig2}
 \end{figure}

Of course, Eq. (\ref{ds2bo}) fails whenever the separation of length scales ansatz is no longer valid, near the points where $\cos(kx)=0$.   In this cross-over region, (shifting $x$ by $-\pi/2k$), we have \begin{equation}
\lim_{z \to 0}A_\mu \mathrm{d} x^\mu \approx Vkx \mathrm{d}t.
\end{equation}This turns out to be in many ways analogous to the $\mathrm{AdS}_4$-RN solution with a background magnetic field, although of course the ``magnetic field" lies in the $xt$ plane, and is thus an electric field on the boundary.      Mathematically, a solution obeying the UV boundary conditions, and neglecting higher order terms in the expansion of $\sin(kx)$, exists:  \begin{equation}
\mathrm{d}s^2_{\mathrm{CO}} = \frac{1}{z^2} \left[\frac{\mathrm{d}z^2}{u(z)} - \mathrm{d}t^2 + \mathrm{d}x^2 + u(z)\mathrm{d}y^2\right]
\end{equation}where \begin{equation}
u(z) = 1+c(Vk)^{3/2}z^3 - V^2k^2 z^4.
\end{equation}Here $c$ is an undetermined O(1) constant which is not important for our aims here, which are strictly qualitative.   The key thing that we note about this metric is that it becomes \emph{strongly anisotropic}.    Furthermore, in the ``magnetic" cross-over regions, the only length scale is $1/\sqrt{Vk}$.   Although the parameter $c$ is unknown and so we cannot do as thorough a comparison as before, we can still check that quantitatively the anisotropy in the crossover regimes scales as $V^{3/2}$ as we predict.   This scaling law follows directly from the fact that the leading order term in $u(z)$ contributes to the boundary expectation values of $\langle T_{xx}\rangle$ and $\langle T_{yy}\rangle$, and $u$ can only depend on the combination $\sqrt{Vk}$.   Furthermore, note from Eq. (\ref{eq:holorenorm}) that $\langle T_{xx}-T_{yy}\rangle$ is proportional to this leading order coefficient in $u$ in the cross-over regions.  We find numerically that this scaling is quantitatively obeyed by the time we reach $V/k=10$ as we show in Fig. \ref{BOfig2}.

Next, let us justify more carefully our argument that Eq. (\ref{ds2bo}) is valid away from the cross-over regions, and precisely breaks down in these cross-over regions.    Following earlier work from the fluid-gravity correspondence \cite{bmw}, we have chosen to make a Born-Oppenheimer ansatz which does not contain any coordinate singularities at the ``local horizon", so that we remove artifact coordinate singularities in the gradient expansion.   In this gauge, we can compute \begin{equation}
\Delta^{\mu\nu} \equiv \left[G^{\mu\nu}[\mu(x)] - 3g^{\mu\nu}[\mu(x)] - \frac{1}{2}T^{\mu\nu}[\mu(x)]\right]
\end{equation}Since obviously if we plug in the Born-Oppenheimer metric with a constant $\mu(x)$, the expression for $\Delta^{\mu\nu}=0$, we see that $\Delta^{\mu\nu}$ will capture the deviations from Einstein's equations.    So long as $|\Delta^{\mu\nu}|$ is much smaller than either $|G^{\mu\nu}|$ or $|T^{\mu\nu}|$,\footnote{Einstein's equations for us read $G^{\mu\nu} - 3g^{\mu\nu} - T^{\mu\nu}/2 = 0$.   At least two of these terms must \emph{always} be ``large" and of the same order to satisfy the equations.   Thus, we can always ignore $g^{\mu\nu}$ since either $G^{\mu\nu}$ or $T^{\mu\nu}$ must be large.} we trust the Born-Oppenheimer approximation, since the equations of motion are perturbatively satisfied.   In fact, for Eddington-Finkelstein coordinates, there is only one non-zero component of this tensor: \begin{equation}
\Delta^{rr} = k^2V^2 z^6 \frac{24\sin^2(kx)+2V^2z^2\cos^2(kx) - 4\sqrt{3}Vz\cos(kx)(1+\sin^2(kx))}{48}.
\end{equation}For comparison, let us look at a ``typical" term in the Einstein equations, so that we may compare the sizes of the two terms at various $(z,x)$.      One can easily check that \emph{all} tensor components in $g^{\mu\nu}$, $G^{\mu\nu}$, and $T^{\mu\nu}$, evaluated on the Born-Oppenheimer metric \emph{before} taking spatial derivatives, are of the form \begin{equation}
G^{\mu\nu}, g^{\mu\nu}, T^{\mu\nu} \sim z^2 P\left(\frac{z}{z_+(x)}\right)
\end{equation}where $P$ is some polynomial function -- of course, the generic form expression also follows directly from dimensional grounds, although explicit analysis shows us that $P$ is always finite.   As much as we can below, we will absorb all factors of $Vz$ into a $z/z_+$, as this  ratio is bounded and the Born-Oppenheimer geometry is dependent only on this ratio.  Note that not all polynomials $P$ vanish at the local RN horizons, so in order for the spatial modulation to have an important effect, we must therefore have $\Delta^{rr} \sim z^2$.   We find that spatial modulation becomes a nonperturbative correction when \begin{equation}
\frac{\Delta^{rr}}{z^2} \sim V^2k^2z^4C\left(\frac{z}{z_+},x\right).
\end{equation}Here $C(x, z/z_+)$ is some $x$-dependent bounded function which only depends on the ratio $z/z_+$.   Thus, we require that $z \gg 1/\sqrt{Vk}$ for nonperturbative effects to kick in -- this is allowed at $\mathrm{O}(\sqrt{k/V})$ points in $x$ near the turning points, 
corresponding precisely to the crossover regions.

Let us also look at Maxwell's equations.     We find that the non-zero components of $\nabla_\mu F^{\mu\nu}$ are given by \begin{equation}
\nabla_\mu F^{\mu x} \sim \frac{z}{z_+} kVz^3 \sin(kx), \;\;\; \nabla_\mu F^{\mu z} \sim k^2z^3 \left(V^2z^2+ \frac{z}{z_+} + \frac{z^2}{z_+^2}\right)
\end{equation}In the latter equation we are neglecting the O(1) coefficients in the parentheses.   For comparison, the non-zero ``contributions" to Maxwell's equations for the Born-Oppenheimer metric come solely from $\nabla_\mu F^{\mu t} \sim \partial_z F^{zt} \sim z (z/z_+)^2$.    We see that once again, the corrections kick in at $z\sim 1/\sqrt{Vk}$.

We stress that it was crucial to work in a metric without coordinate singularities for the above argument to be valid.   Working with a Born-Oppenheimer ansatz in the more typical Fefferman-Graham coordinates, one will find that the gradient expansion breaks down everywhere deep in the IR, despite the fact that they are to leading order the same metric.   When we promote $\mu(x)$ to have spatial dependence, depending on the gauge in which we write the metric, certain $\mathrm{O}(k/\mu)$ terms will be included or suppressed.     Coordinate singularities can amplify such terms deep in the IR and promote them to non-perturbative corrections, which is why we chose Eddington-Finkelstein coordinates.

\begin{figure}[H] 
   \centering
      \includegraphics[width=4.5in]{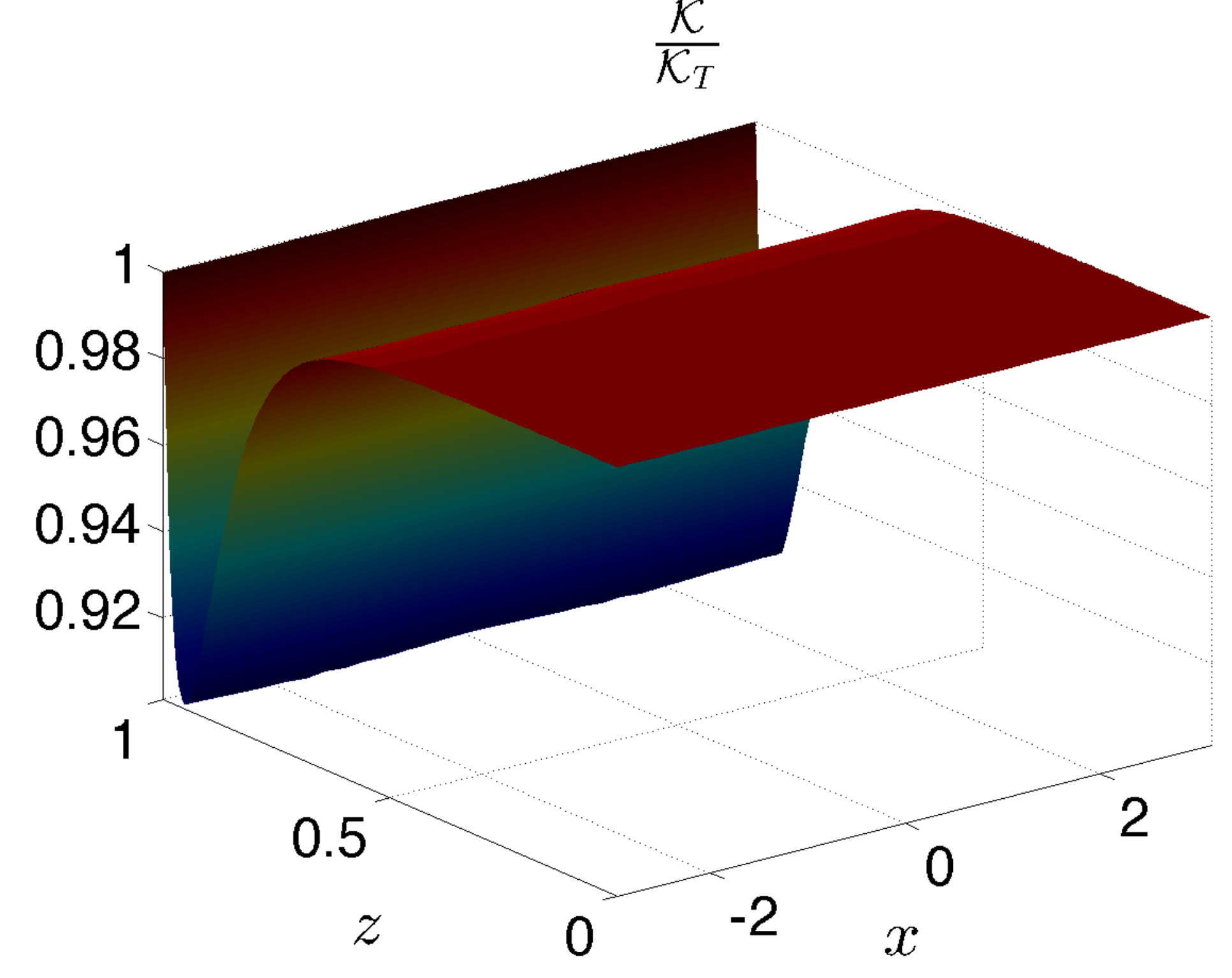}
         \caption{A plot of the invariant $\mathcal{K}/\mathcal{K}_T$ at $T=0.0385$ for $V=1$ and $k=1$.}
   \label{kktplot1}
 \end{figure}

\begin{figure}[H] 
   \centering
      \includegraphics[width=4.5in]{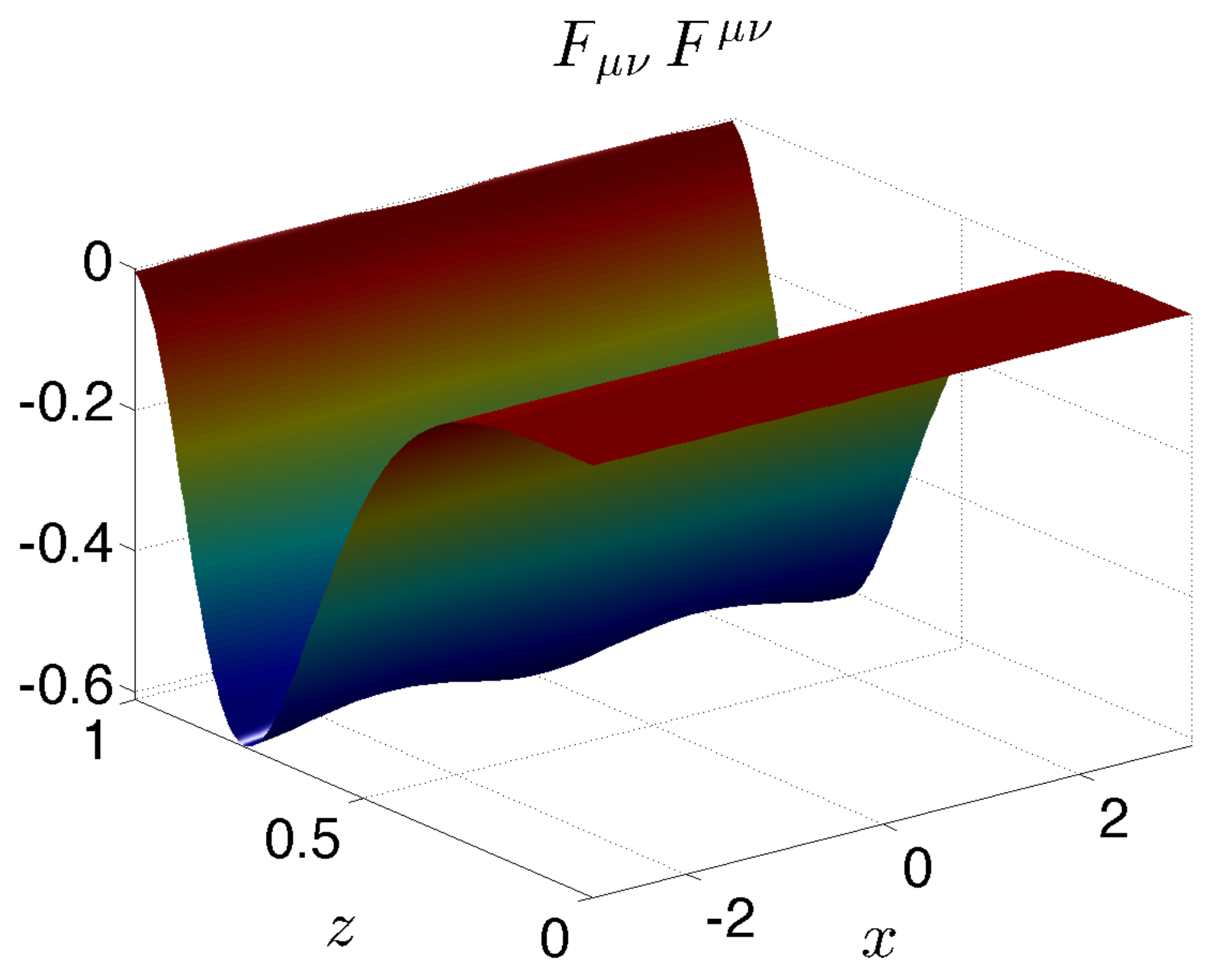}
   \caption{A plot of the invariant $F_{\mu\nu}F^{\mu\nu}$ at $T=0.0385$ for $V=1$ and $k=1$.}
   \label{kktplot2}
\end{figure}

 Let us conclude our discussion of the non-perturbative geometry with some speculation on the IR geometry.   In our numerics, we imposed the requirement that the periodic potential is an irrelevant deformation in the IR.    At next-to-leading order in the gradient expansion, this is a different boundary condition than the boundary conditions of our Born-Oppenheimer metric above.   We should keep in mind that the Einstein-Maxwell system is nonlinear and may admit many interesting classes of solutions, some even with identical boundary conditions, so this should not be seen as problematic.   For the geometry we studied numerically, as we saw in Figure \ref{fig:sigmaholV}, the d.c. conductivity is very close to $\sigma_\infty$, the value it would take if the geometry was pure AdS in the IR, with only a small relative rescaling, even for $V\gg k$.   We are led to speculate that the geometry returns to $\mathrm{AdS}_4$ in the IR, even when $V\gg k$, in the regular solution to Einstein's equations with our UV boundary conditions, at $T=0$.   This means that the periodic potential is an irrelevant operator in the IR, and the low energy effective theory is again conformal, just as in the non-perturbative case.   Geometrically, this is the statement that the Born-Oppenheimer ansatz breaks down at all spatial points $x$, far enough in the bulk, due to corrections caused by the break down of the gradient expansion.   At $V\sim k$, we are able to perform a strong check of this claim by computing geometric invariants, such as the Kretschmann scalar\begin{equation}
\mathcal{K} = R_{\mu\nu\rho\sigma} R^{\mu\nu\rho\sigma}.
\end{equation}For the AdS black brane at temperature $T$, in our numerical gauge, this takes the simple form \begin{equation}
\mathcal{K}_T = 24 + 12\left(\frac{z}{3/4\pi T + (1-3/4\pi T)z}\right)^3.
\end{equation}As we show in Figs. \ref{kktplot1} and \ref{kktplot2}, we can use these invariants to study the geometry in the deep IR in a gauge-invariant way.    The fact that $\mathcal{K} \rightarrow \mathcal{K}_T$ in the deep IR beyond the perturbative paradigm of the previous subsection is consistent with the IR geometry being an AdS black hole geometry, with the same AdS radius as the UV geometry.   Note that the black brane is a consequence of the finite temperature numerics; this argument should hold over at $T=0$.   Note that most of the structure in these plots is located near $z=1$.   This is a consequence of our gauge choice squeezing many points close to $z=1$, and should not be interpreted as physical:  momentum-carrying perturbations decay exponentially in an IR-asymptotically AdS space.   We also show the behavior of $F_{\mu\nu}F^{\mu\nu}$, a geometric invariant which shows us that, as expected, the gauge fields die off deep in the IR.    Again, we have not proven that the deep IR physics is conformal and described by an AdS geometry, but we believe this is a reasonable conjecture.

\subsubsection{Conductivity}

\begin{figure}[H] 
   \centering
   \includegraphics{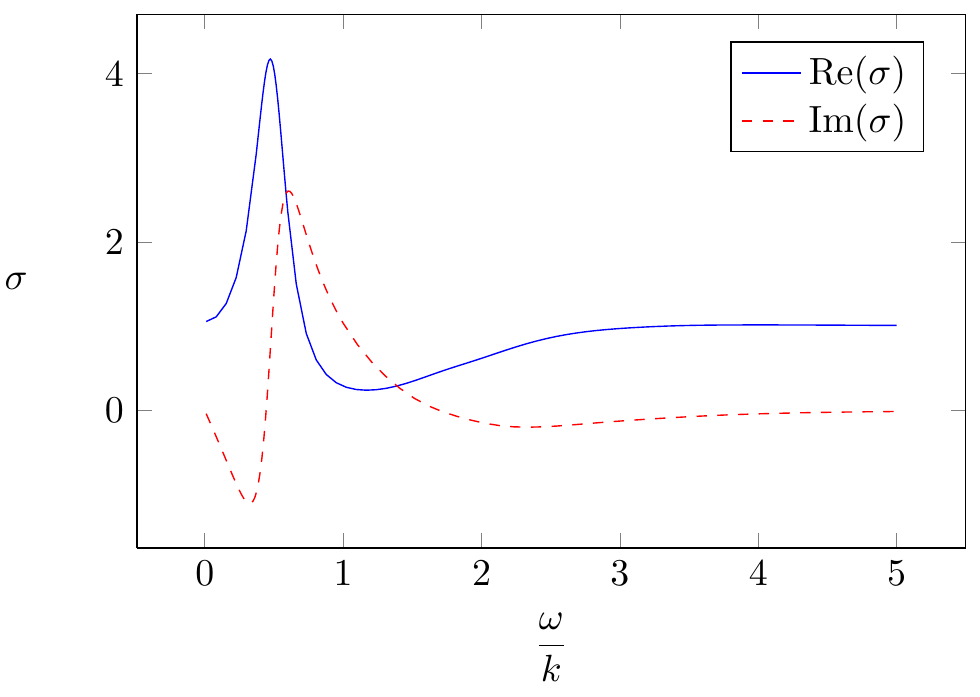}
   \caption{A plot of $\sigma$ vs. $\omega$ for $V/k=4$, taken at $k/T=8$.  }
   \label{fig:sigmahol2}
\end{figure}

Now, let us ask what happens to the conductivity $\sigma(\omega)$ in the nonperturbative regime.   Our numerical results are shown in Figure \ref{fig:sigmahol2}.   We see a couple dominant features.   Firstly, the d.c. conductivity is \emph{very close} to $\sigma_\infty$, as we have already seen in Figure \ref{fig:sigmaholV}.   The $\delta$ function peak in the conductivity of the RN black hole has been smoothed out, and most of the spectral weight lies at $\omega \sim k$, as we will discuss more shortly.   We then transition into a regime where the conductivity dips below $\sigma_\infty$.  At this point, the conductivity is qualitatively approximated by the conductivity of a RN black hole, corresponding to a CFT with chemical potential $V/\sqrt{2}$.   For $\omega \gg V$, the conductivity returns to $\sigma_\infty$, as we expect.

We note that the study of Ref.~\cite{gary2} also found a `resonance' peak in the conductivity at $\omega \sim k$, but this appeared
on top of stronger `Drude' peak at $\omega=0$; the latter appears because they had $\mu_0 \neq 0$.

  To understand the features of the conductivity from first principles, let us begin by assuming that the Born-Oppenheimer approximation is valid at the scale we are studying.    We then can approximate that the conductivity is given by Eq. (\ref{boeq1}), which states that $\sigma(\omega) \sim \int \mathrm{d}x \sigma(\omega,x)$ where $\sigma(\omega,x)$ is the local conductivity associated to the local RN ``black hole" at spatial point $x$.   This is a valid approximation because the probe field dies off very quickly long before the effects of the background gauge field alter the geometry significantly.   Since the RN geometry has only a single energy scale $\mu$, we conclude that \begin{equation}
\sigma(\omega,x) = \Phi\left(\frac{\omega}{V|\cos(kx)|}\right),\label{boeq2}
\end{equation}where $\Phi$ is a universal function whose analytic form is unknown (plots of this function can be found in \cite{hartnollcqg}, \emph{e.g.}).  Combining Eqs. (\ref{boeq1}) and (\ref{boeq2}) we find \begin{equation}
\sigma(\omega)\approx 2 \int\limits_0^1\frac{\mathrm{d}\zeta}{\sqrt{1-\zeta^2}} \; \Phi\left(\frac{\omega}{V\zeta}\right).   \label{boeq3}
\end{equation}For $\omega \sim V$, this formula suggests that the conductivity should roughly be RN conductivity, where a dip begins to appear, an effect which we see in Figure \ref{fig:sigmahol2}.   Of course, $\Phi(x) \sim \delta(x)$ for small $x$, and so this approximation must break down at some point. 

We can use this argument to connect the scales at which new physics appears in the conductivity to the scales at which spatial modulation affects the geometry in an important way.   Let us consider the equation of motion for the gauge field $A_x$ more carefully.   It is this equation whose boundary behavior determines the universal function $\Phi$.   We're going to compute the conductivity by considering the equation of motion where we treat $\mu$ as a constant.   Although obviously this approximation is not appropriate for ``small" frequencies, our goal is to determine how the geometry determines what we mean by the word ``small".    If the function $A_x$ is concentrated outside a regime where spatial modulation alters the Born-Oppenheimer ansatz, then we expect RN conductivity to be a good approximation;  otherwise, we conclude that the striping has induced new behavior in the conductivity.   The equation of motion neglecting the striping is \begin{equation}
\partial_z \left(f(z)\partial_z A_x\right) = - \frac{\omega^2}{f(z)}A_x + 4\mu^2 \left(\frac{z}{z_+}\right)^2 A_x,
\end{equation}
where $\mu$ is the local chemical potential at the $x$ co-ordinate under consideration.    Since we are looking at near horizon physics, let us define $w = \sqrt{12}-z\mu$ to be a new co-ordinate.
Switching coordinates to $u=1/w$, we can show that up to O(1) factors, for large $u$ the gauge field equation of motion reduces to\begin{equation}
\partial_u^2 A_x = \left(\frac{1}{u^2}- \frac{\omega^2}{\mu^2}\right)A_x
\end{equation}Note this relies on $f(w)\sim w^2$ for small $w$.   By  rescaling to $u = \mu U /\omega$:  \begin{equation}
\partial_U^2 A_x = \left(\frac{1}{U^2} - 1\right)A_x.
\end{equation}So the behavior of the function $A_x$ at low frequencies should be universal and only depend on $\omega/\mu w$, \emph{if} the Born-Oppenheimer geometry holds.

\begin{figure}[here] 
   \centering
   \includegraphics{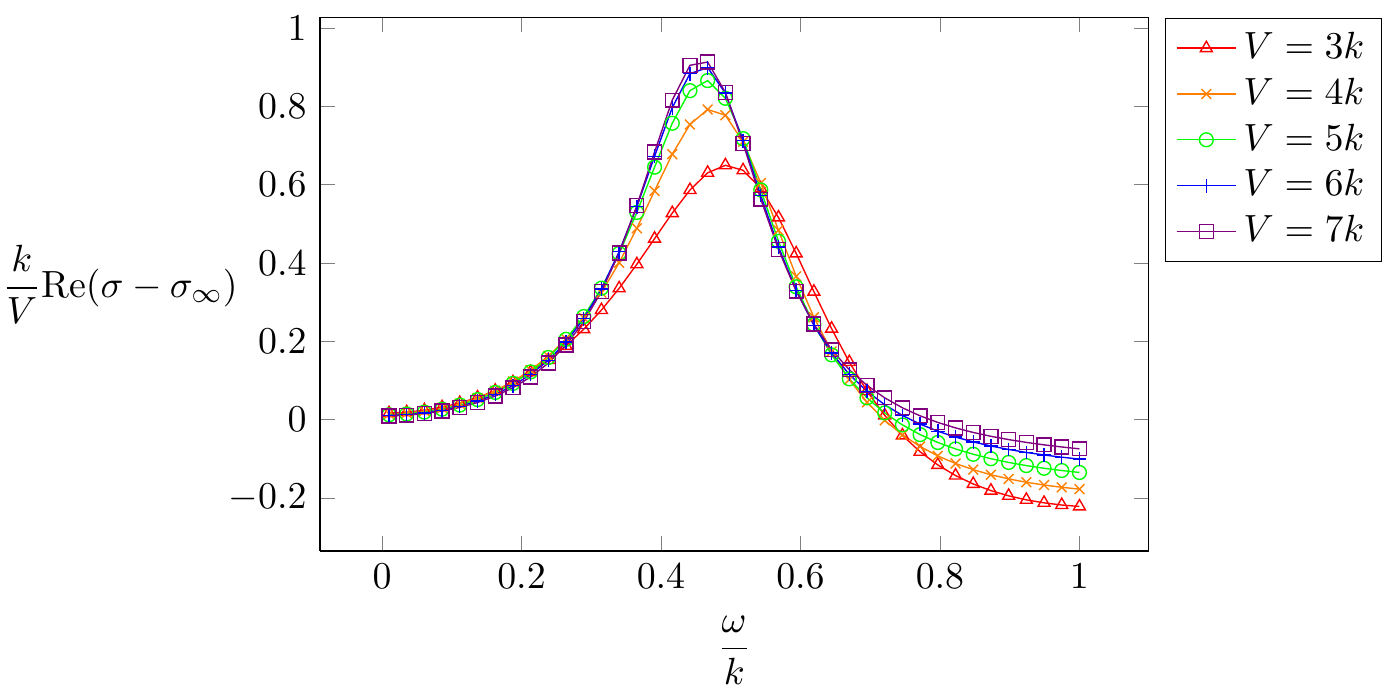} 
   \caption{This figure shows that for $\mu \gg k$, $\mathrm{Re}(\sigma)$ approaches a universal function $\times V /k$ for $\omega<k$.   The deviations for $\omega \sim k$ demonstrate the transition into an intermediate regime which transitions between the peak and the RN conductivity.   This data was taken at $k/T=8$.}
   \label{varymu}
\end{figure}

By far the dominant effect which is observed numerically is a strong peak which emerges at $\omega \sim k$.      Remarkably, the peak in $\sigma(\omega)$ at $\omega \sim k$ is described by $V/k$ times a universal function once $V\gg k$, as we show in Figure \ref{varymu}.    This qualitative scaling in the conductivity is quite easy to understand heuristically using a sum rule:\begin{equation}
\int\limits_0^\infty \mathrm{d}\omega \; \mathrm{Re}(\sigma(\omega)-\sigma_\infty) = 0.
\end{equation}We can break this integral into two pieces:  one for $\omega<k$, and one for $k<\omega<V$.    The former regime is dominated by the large peak at finite $\omega$, and the latter regime is dominated by the large dip in the conductivity associated with the RN conductivity.   We then have, heuristically: \begin{equation}
\sigma_{\mathrm{max}} k + (-\sigma_{\infty})V=0,
\end{equation}which implies that $\sigma_{\mathrm{max}}$, the maximal value of $\mathrm{Re}(\sigma)$, should scale as \begin{equation}
\sigma_{\mathrm{max}} \sim \frac{V}{k}\sigma_\infty.
\end{equation}
Geometrically, this peak can be interpreted as a quasinormal mode of the black hole.   In the field theory, we suspect that this is a consequence of broken translational symmetry and the presence of low-lying excitations at a finite momentum.

Thus, if the conductivity experiences new physics when $\omega \sim k$, it would suggest that the approximation of Eq. (\ref{boeq3}) has broken down, at all points $x$, at a small distance $w\sim k/V$ from the horizon.    Of course, perturbation theory may still be quite good at describing the background geometry, but the first order perturbation might lead to nonperturbative corrections to transport functions such as optical conductivity, because turning on this perturbation couples $A_x$ to new graviton modes.   Further analytic exploration of transport in weakly inhomogeneous systems is worthwhile to resolve this issue.

\section{Weakly-coupled CFTs}
\label{sec:graphene}

As discussed in Section~\ref{sec:intro}, a convenient paradigm for a weakly-coupled CFT is a theory of $N_f$ Dirac fermions $\psi_\alpha$ coupled to a SU($N_c$) gauge field $a_\mu$ with Lagrangian as in Eq.~(\ref{L0}).
To this theory we apply a periodic chemical potential which couples to the globally conserved U(1) charge
\beq
\mathcal{L}_V = - V \cos (k x) \mathrm{i} \sum_{\alpha=1}^{N_f} \overline{\psi}_\alpha \gamma^0 \psi_\alpha . \label{LV}
\eeq

\begin{figure}[h]
\begin{center}
\includegraphics[width=4.in]{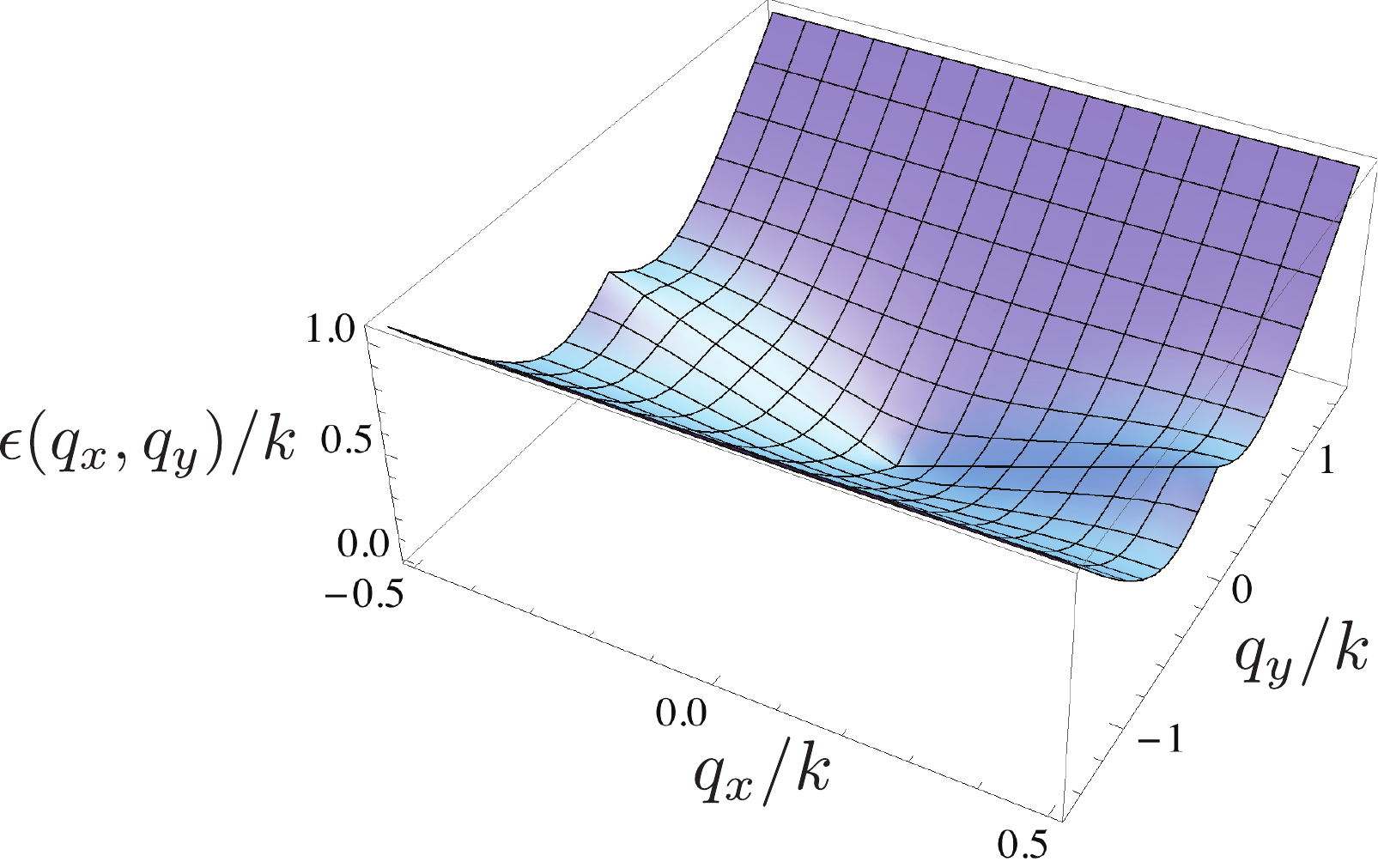}
\caption{Plot of the lowest positive energy eigenvalues $\epsilon(q_x,q_y)$ for $V/k = 0.8$. There is a single Dirac node at $(0,0)$.
The dispersion is periodic as a function of $q_x$ with period $k$, and a full single-period is shown. There is no periodicity as a function of $q_y$, and the energy increases as $|q_y|$ for large $|q_y|$.}
\label{fig:dirac08}
\end{center}
\end{figure}

\begin{figure}[h]
\begin{center}
\includegraphics[width=4.in]{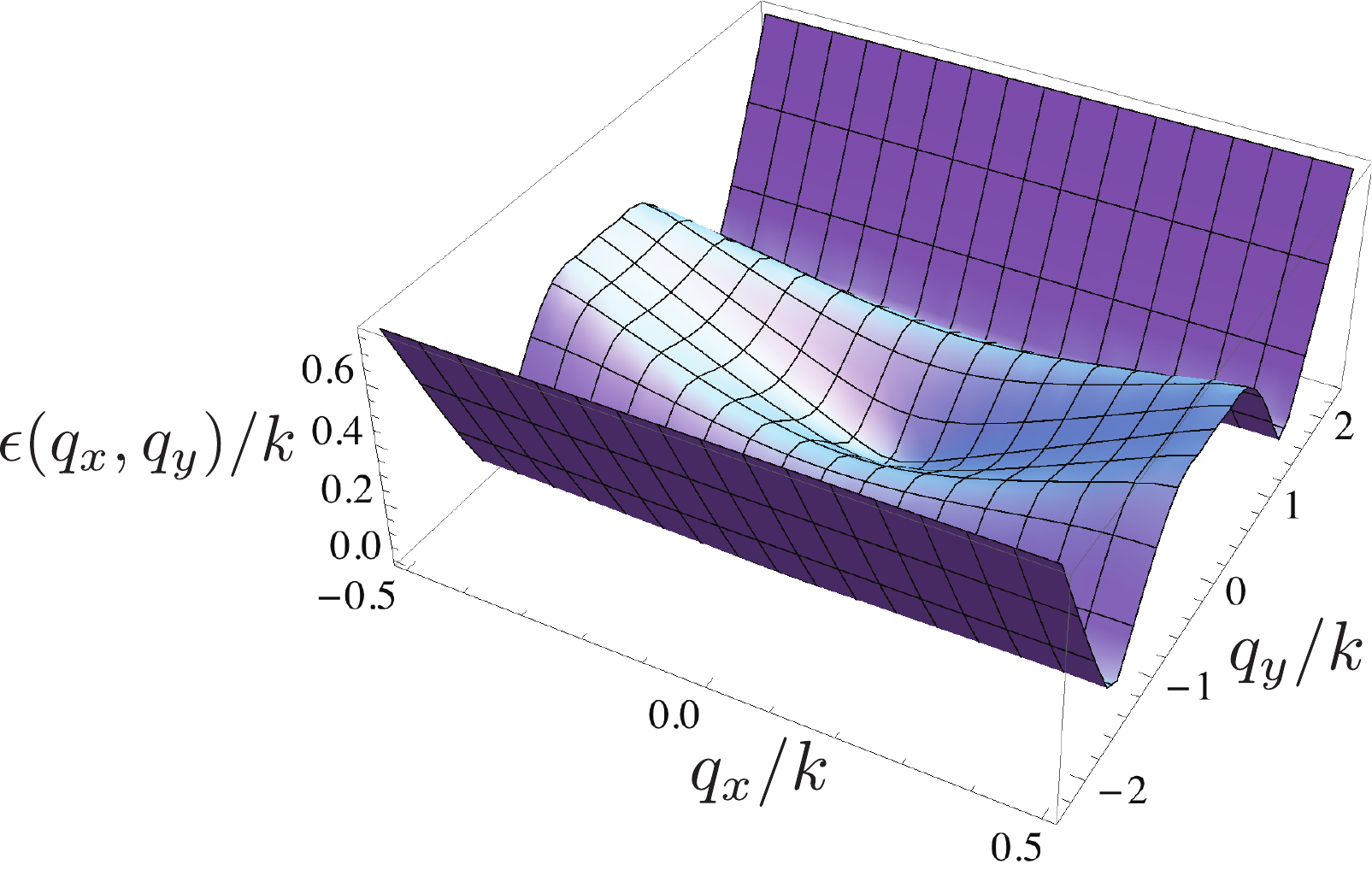}
\caption{As in Fig.~\ref{fig:dirac08}, but for $V/k = 2.0$. Now there are 3 Dirac nodes: one at $(0,0)$,
and a pair at $(0,\pm 1.38)$. One of the latter pair is more clearly visible in Fig.~\ref{fig:diracspec}B.}
\label{fig:dirac20}
\end{center}
\end{figure}

The essential structure of the influence of the periodic potential is clear from a careful examination of the spectrum of the free fermion limit. By Bloch's theorem, the fermion dispersion $\omega = \epsilon (q_x, q_y)$ is a periodic function of $q_x$ with period $k$:
\beq
\epsilon (q_x + k, q_y) = \epsilon (q_x).
\eeq
So we can limit consideration to the ``first Brillouin zone'' $-k/2 \leq q_x \leq k/2$.
This spectrum has been computed in a number of recent works in the context of applications to 
graphene \cite{park1,ho1,park2,park3,brey1,barbier1,wang1,brey2}.
We show results of our numerical computations in Figs.~\ref{fig:dirac08}, \ref{fig:dirac20},  and \ref{fig:diracspec}, 
obtained via diagonalization of the Dirac Hamiltonian
in momentum space (see Appendix~\ref{app:dirac} for details).
 The periodic potential couples together momenta, $(q_x + \ell k, q_y)$ with different integers $\ell$, and we 
 numerically diagonalized the resulting  matrix for each $(q_x, q_y)$.

For small $V/k$, the spectrum can be understood perturbatively. There is a Dirac cone centered at $\vec{q} = (0,0)$ and this undergoes
Bragg reflection across the Bragg planes at $q_x/k = \pm 1/2$, resulting in band gaps at the Brillouin zone boundary. See Figs.~\ref{fig:dirac08} and \ref{fig:diracspec}(A).

However, the evolution at larger $V/k$ is interesting and non-trivial. The spectrum undergoes an infinite set of quantum phase transitions
at a discrete set of values of $V/k$, associated with the appearance of additional Dirac nodes along the $q_y$ 
axis  \cite{park1,ho1,park2,park3,brey1,barbier1,wang1,brey2}. The first of these occurs at phase transitions occurs at 
$V/k = 1.20241..$ \cite{brey1}; for $V/k$ just above this critical value 2 additional Dirac nodes develop along the $q_y$ axis,
and move, in opposite directions, away from $q_y=0$ with increasing $V/k$. This is illustrated by the fermion spectrum at $V/k=2.0$ which
is displayed in Figs.~\ref{fig:dirac20} and~\ref{fig:diracspec}(B).
\begin{figure}[h]
\begin{center}
\includegraphics[width=5.0in]{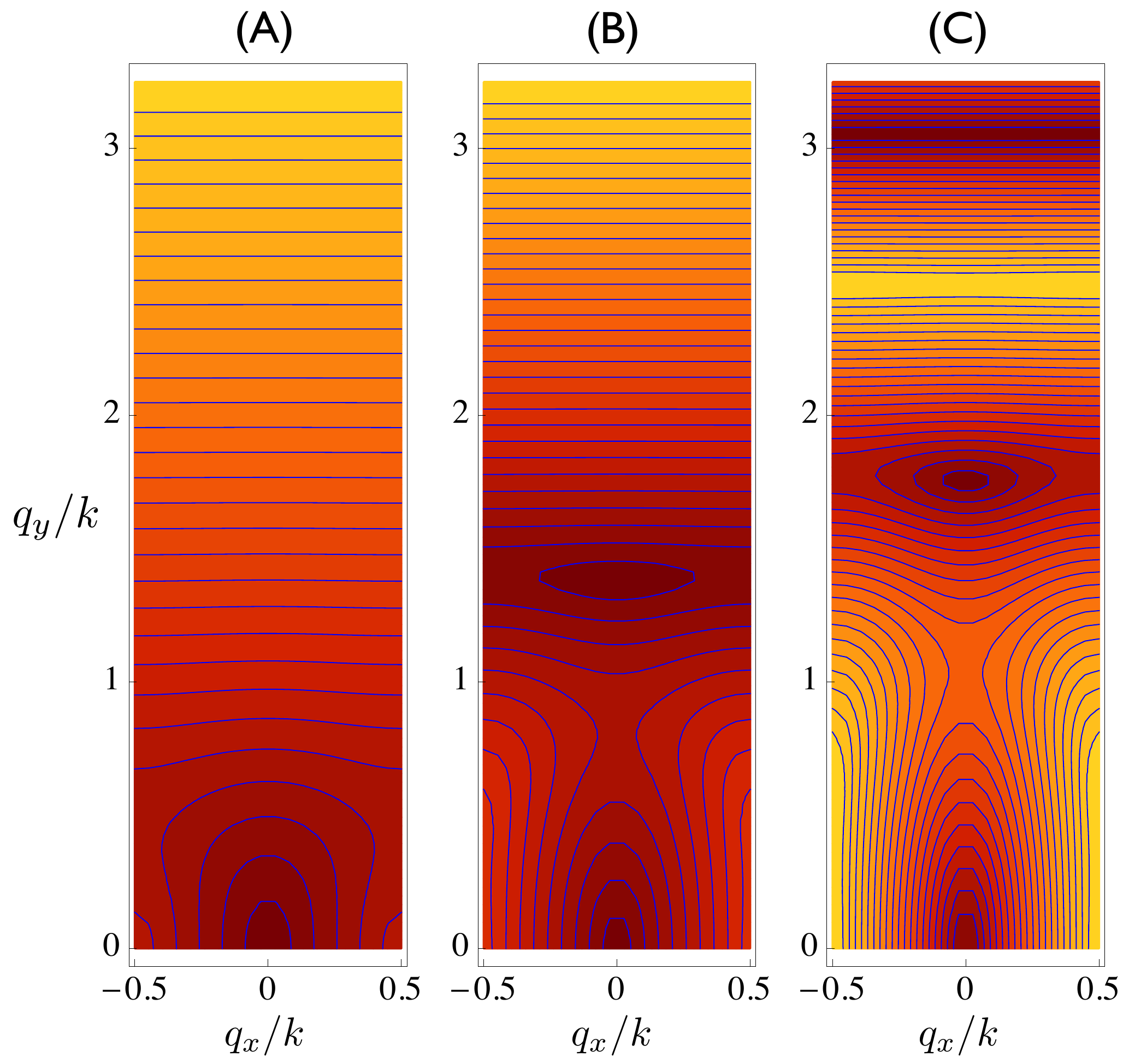}
\caption{Contour plot of the lowest positive energy eigenvalues $\epsilon(q_x,q_y)$ for (A) $V/k = 0.8$, (B) $V/k=2.0$, and (C) $V/k= 3.6$.
All three plots show Dirac nodes at $(0,0)$. However for larger $V/k$, additional Dirac nodes appear at
(B) $(q_x/k=0, q_y/k = \pm 1.38 )$, and (C) $(q_x/k=0, q_y/k = \pm 1.75 )$,  $(q_x/k=0, q_y/k = \pm 3.06 )$ }
\label{fig:diracspec}
\end{center}
\end{figure}

Additional phase transitions appear at larger $V/k$, each associated with an additional pair of Dirac nodes emerging from $q_y=0$.
The second transition is at $V/k = 2.76004..$ \cite{brey1}. This is illustrated by the fermion spectrum at $V/k=4.0$ which
is displayed in Fig.~\ref{fig:diracspec}(C), which has 5 Dirac nodes. Subsequent phase transitions appear at $V/k = \mathcal{J}_n/2$, where
$\mathcal{J}_n$ is the $n$'th zero of the Bessel function $J_0$ \cite{brey1}.

\begin{figure}[H]
\begin{center}
\includegraphics[width=5in]{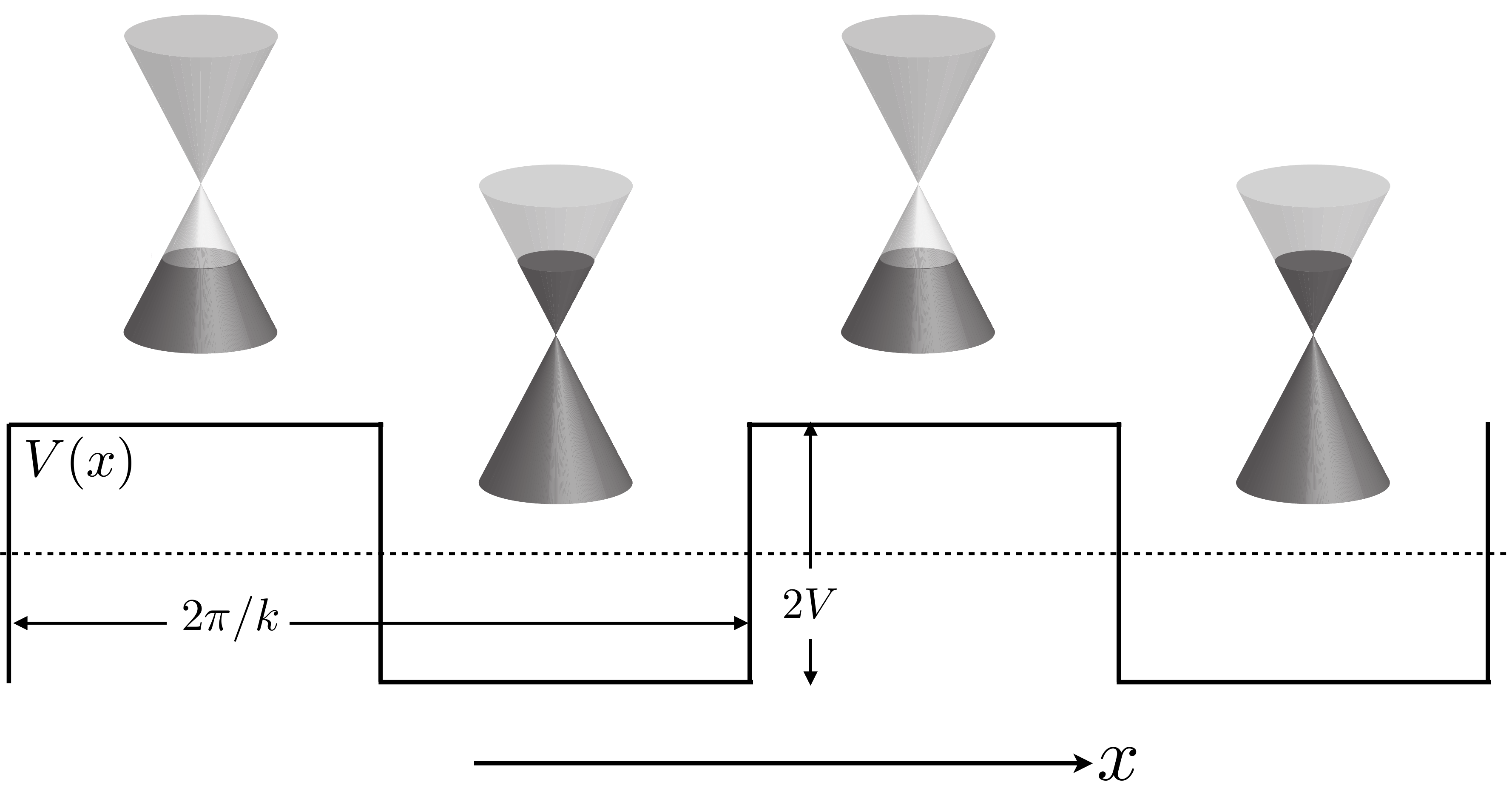}
\caption{Dirac fermions in a periodic rectangular-wave chemical potential. The regions alternate between local electron and hole Fermi surfaces. }
\label{fig:squarewell}
\end{center}
\end{figure}
\begin{figure}[H]
\begin{center}
\includegraphics[width=4in]{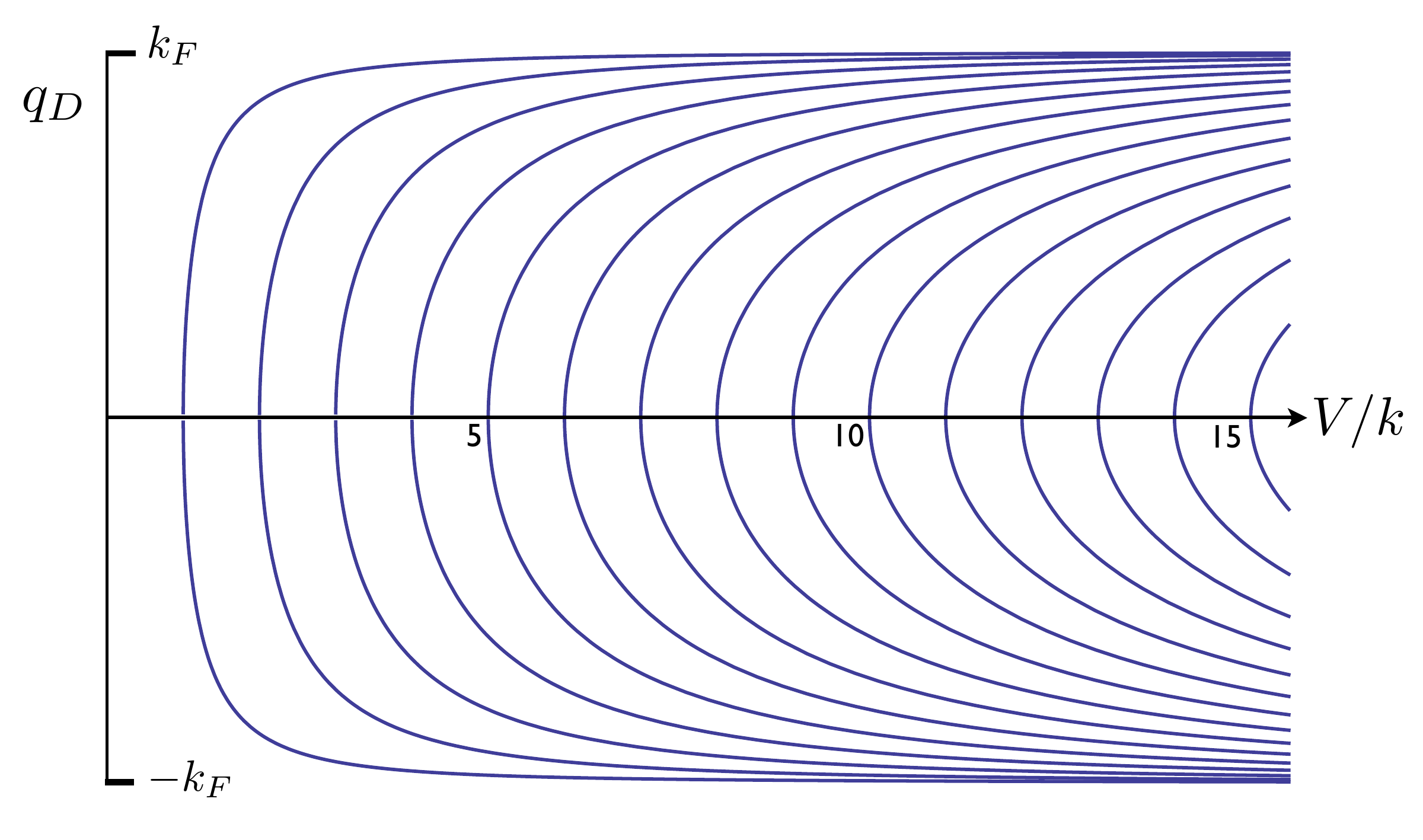}
\caption{Dirac points of the periodic rectangular-wave potential in Fig.~\ref{fig:squarewell} obtained from Eq.~(\ref{qD}).
The Dirac points are at $(0,q_D)$ and move as a function of $V/k$ as shown. Two new zero energy Dirac points emerge at each integer $V/k$.
There is also a Dirac point at $(0,0)$ for all $V/k$.}
\label{fig:diracpts}
\end{center}
\end{figure}
Some physical insight into these additional Dirac nodes can be obtained by considering instead a simpler 
periodic rectangular-wave potential, illustrated in Fig.~\ref{fig:squarewell}.  The chemical potential is now piecewise constant, and in each region there are electron-like or hole-like Fermi surfaces of radius $k_F = V$.
It is a simple matter to include tunneling between these regions as described in Appendix~\ref{app:zeros}, 
and the resulting spectrum is qualitatively similar to Fig.~\ref{fig:diracspec},
with an infinite number of quantum transitions associated with the appearance of pairs of Dirac nodes. However, it is now possible to write down a
simple expression for the positions of these nodes \cite{barbier1,wang1}: the nodes are at $q_x = 0$ and $q_y = \pm q_D (n)$ where
\beq
q_D (n) = \pm k_F \sqrt{ 1 - n^2 k^2 /V^2}, \label{qD}
\eeq
and $n$ is a positive integer. Thus the Dirac node at $q_D (n)$ appears for $V/k > n$, and so the $n$'th quantum transition is now at $V/k = n$.
See Fig.~\ref{fig:diracpts}.

\begin{figure}[h]
\begin{center}
\includegraphics[width=4.5in]{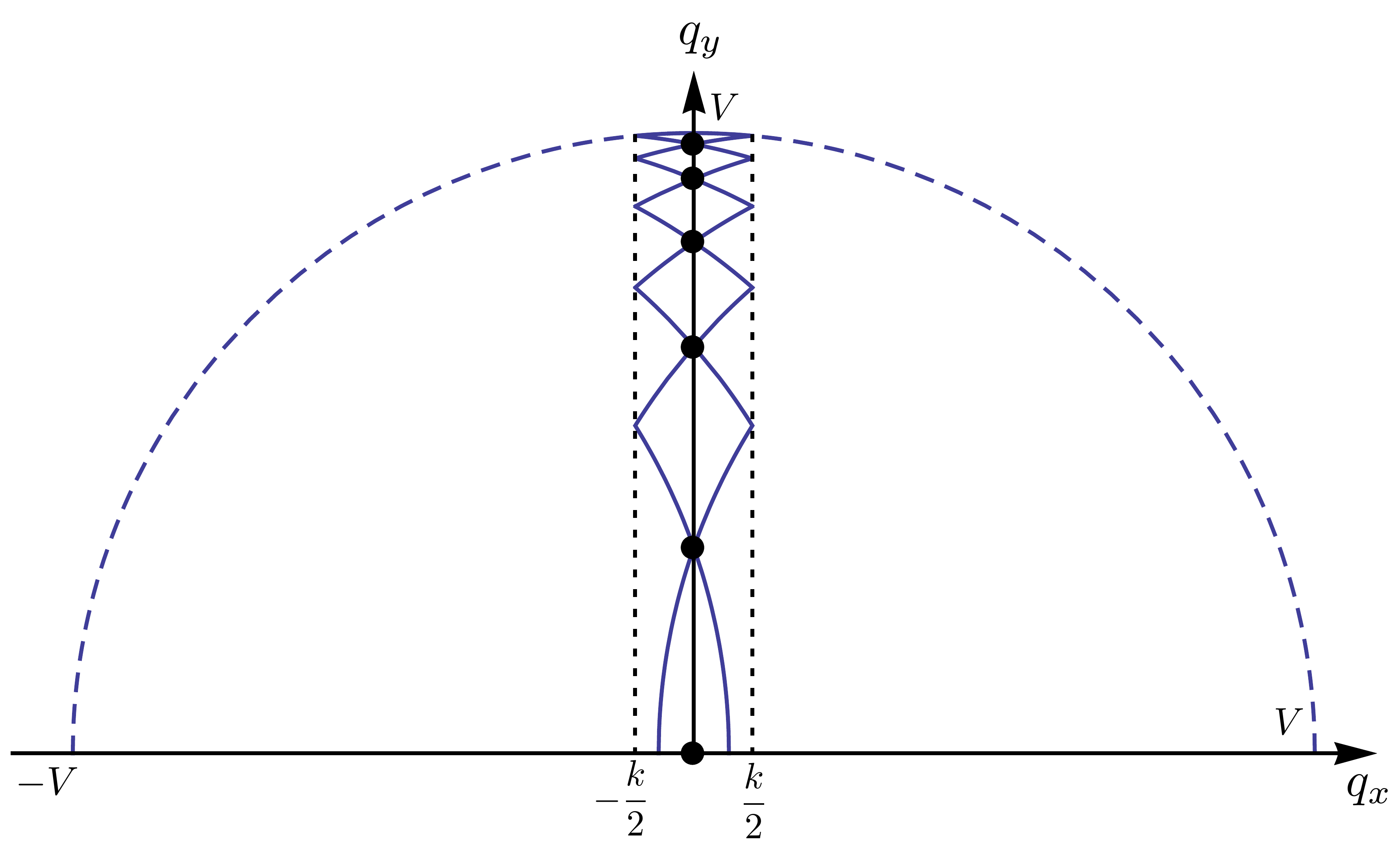}
\caption{Illustration of the positions of the Dirac points with positive $q_D$ for $V/k = 5.3$.
The dashed line is the location of the electron and hole Fermi surfaces of Fig.~\ref{fig:squarewell}.
These are folded back into the first Brillouin zone $-k/2 < q_x < k/2$ and shown as the full lines.
The Dirac points are the filled circles at the positions in Eq.~(\ref{qD}), and these appear precisely at the intersection points of the 
folded Fermi surfaces in the first Brillouin zone.
}
\label{fig:diracpts2}
\end{center}
\end{figure}

A graphical derivation of the positions of the Dirac points is shown in Fig.~\ref{fig:diracpts2}, and this shows that they appear
precisely at the intersection points of the electron and hole Fermi surfaces of Fig.~\ref{fig:squarewell}, after they have been folded
back into the first Brillouin zone.
Thus we may view these points as the {\em remnants\/} of the local Fermi surfaces obtained in a Born-Oppenheimer picture of the periodic 
potential.
  
We can also use the simpler rectangular-wave model to compute the velocities $v_x (n)$ and $v_y (n)$ characterizing the $n$'th
Dirac node at $(0, q_D (n))$.  Their values are \cite{barbier1} (see Appendix~\ref{app:zeros})
\beq
v_x (n) =1 -  \left( \frac{q_D (n)}{k_F} \right)^2 \quad , \quad v_y (n) = \left( \frac{q_D (n)}{k_F} \right)^2 . \label{velocities}
\eeq
So we see that the Dirac points with small $|q_D (n)|$ move predominantly in the $x$ direction, while those with large $|q_D (n)|$ move
predominantly in the $y$ direction. This is also consistent with our remnant Fermi surface interpretation: the Fermi velocity for small $q_y$
is oriented in the $x$ direction, while the Fermi velocity for $q_y \approx k_F$ is oriented in the $y$ direction. 
Note also that the trends in the velocities for the cosine potential, as deduced from the dispersions in Fig.~\ref{fig:diracspec},
are the same as that of the rectangular-wave potential.

We started here with a theory which had $N_f$ massless Dirac fermions. After applying a periodic potential, we end up with a theory
which is described by $N_D N_f$ Dirac fermions at low energies, 
where $N_D \geq 1$ is an integer. For the periodic rectangular-wave potential 
\beq
N_D = 2 \left\lfloor V/k \right\rfloor + 1.
\eeq
For the cosine potential, $N_D$ is a similar piecewise-constant function of $V/k$, determined by the zeros of the Bessel function \cite{brey1}.
We can now add interactions to the low energy theory: by gauge-invariance, the $a_\mu$ gauge field will couple minimally to each of 
the $N_D N_f$ Dirac fermions, and so the effective theory will have the same structure as the Lagrangian in Eq.~(\ref{L0}). 
A crucial feature of this theory is that the number of massless Dirac fermions is {\em stable\/} to all orders in perturbation theory, and so our
picture of emergent Dirac zeros continues to hold also for the interacting theory. This stability of the Dirac zeros can be viewed as a 
remnant of the Luttinger theorem applied to the parent Fermi surfaces from which the Dirac zeros descend (Fig.~\ref{fig:diracpts2}).

\begin{figure}[H]
\begin{center}
\includegraphics[width=3.5in]{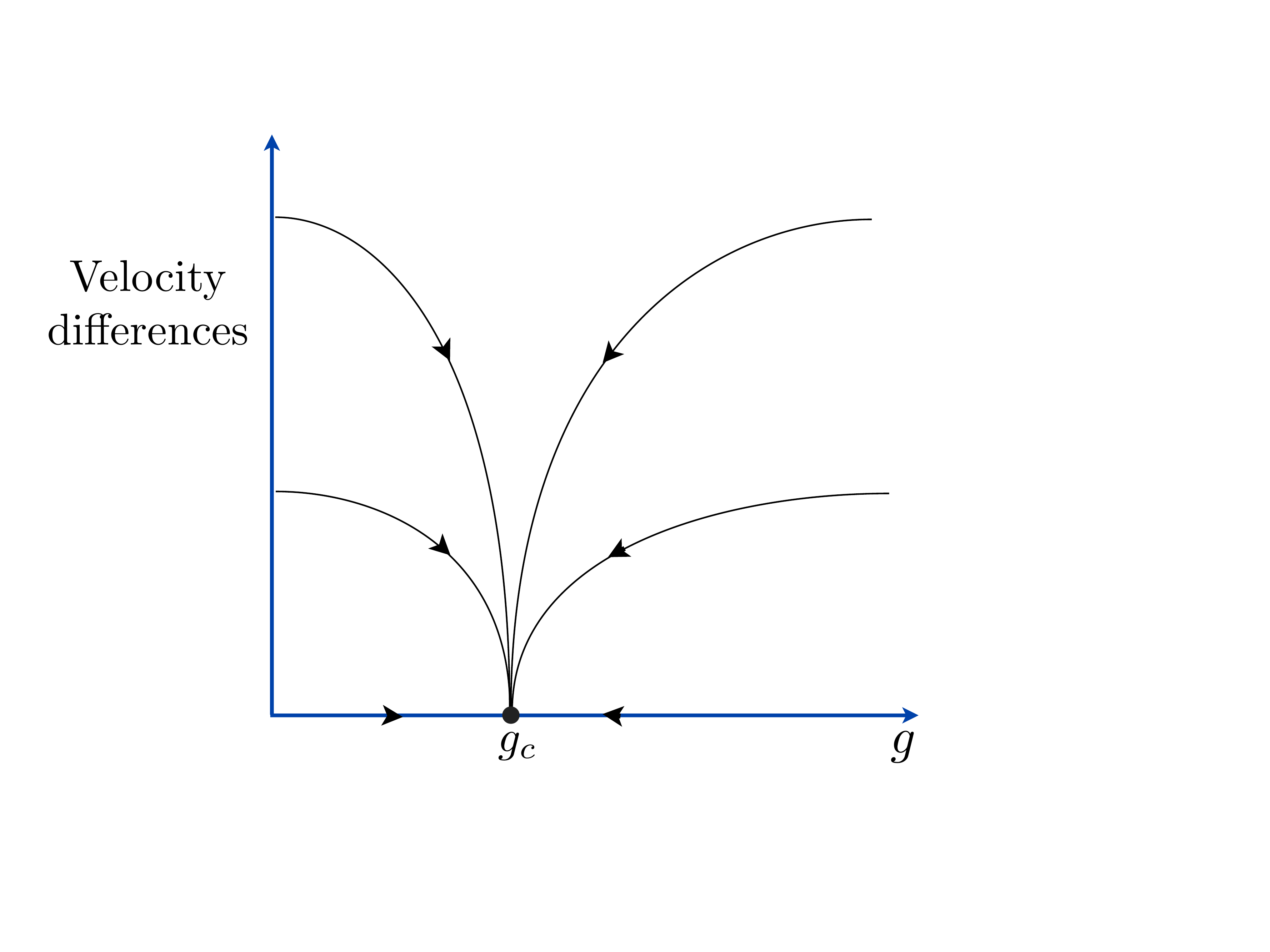}
\caption{Schematic of the expected renormalization group flow (towards the IR) for the field theory in Eq.~(\ref{L0}), 
after we allow for velocity anisotropies, and a bare gauge kinetic energy $f^2 /(4g)$ where $f = \mathrm{d}a$. 
The $1/N_f$ expansion of Ref.~\cite{igor} automatically places the theory at the fixed point $g=g_c$, even in the absence of an explicit
gauge kinetic energy.}
\label{fig:velocities}
\end{center}
\end{figure}

However, this low energy theory of $N_D N_f$ Dirac fermions is not, strictly speaking, a CFT. This is because the velocities in (\ref{velocities})
are a function of $n$, and it not possible to set them all to unity by a common rescaling transformation. However, once we include interactions between the Dirac fermions from the SU($N_c$) gauge field in Eq.~(\ref{L0}), there will be renormalizations to the velocities from 
quantum corrections. As shown in Ref.~\cite{hermele}, such renormalizations are expected to eventually scale all the velocities to a 
common value (see Fig.~\ref{fig:velocities}). 
So the ultimate IR theory is indeed a CFT of $N_D N_f$ Dirac fermions coupled to a SU($N_c$) gauge field. This is distinct from the UV theory,
which had only $N_f$ Dirac fermions coupled to a SU($N_c$) gauge field. 

Finally, we can also consider the transition points between these IR CFTs, which appear at critical values of $V/k$.
Here we find \cite{barbier1} the fermion dispersion 
\beq
\epsilon (0, q_y) \sim |q_y|^3 \quad, \epsilon (q_x, 0) \sim |q_x|.
\label{qcubed}
\eeq
This critical theory is evidently not relativistic, and it would be interesting to compute the effects of interactions at 
such a point.

\subsection{Conductivity}
\label{graphene:cond}

This subsection will compute the frequency dependent conductivity, $\sigma (\omega)$ of free Dirac fermions in the presence of the cosine
potential in Eq.~(\ref{LV}). Previous results in the graphene literature have been limited to the d.c. conductivity $\sigma (0)$. 

We use the standard Kubo formula applied to the band structure computed above, to compute $\sigma (\omega)$ directly at $T=0$ 
(see Appendix~\ref{app:dirac}).
In the absence of a periodic potential, the free Dirac fermions yield the frequency independent result \cite{tasi}
\beq
\sigma (\omega) = \frac{N_f}{16} \quad , \quad V=0. \label{sigmafree}
\eeq
For graphene, this is the conductivity measured in units of $e^2/\hbar$, and we have to take $N_f = 4$.

We begin by examining the frequency-dependence of the conductivity at small $V/k$, in the regime where
there is only a single Dirac node, as in Fig.~\ref{fig:diracspec}A. The results were shown in Fig.~\ref{fig:sigma0p5}.
At large $\omega$, the result approaches the value in Eq.~(\ref{sigmafree}), and this will be the case for all our results below.
At small $\omega$ there is interesting structure in the frequency dependence induced by inter-band transitions in the presence of the 
periodic potential, including at dip
centered at $\omega = k$. Notice the remarkable similarity of this frequency dependence to that obtained in holography
at small $V/k$, as shown in Fig.~\ref{fig:sigmahol1}.

\begin{figure}[H]
\begin{center}
\includegraphics[width=5in]{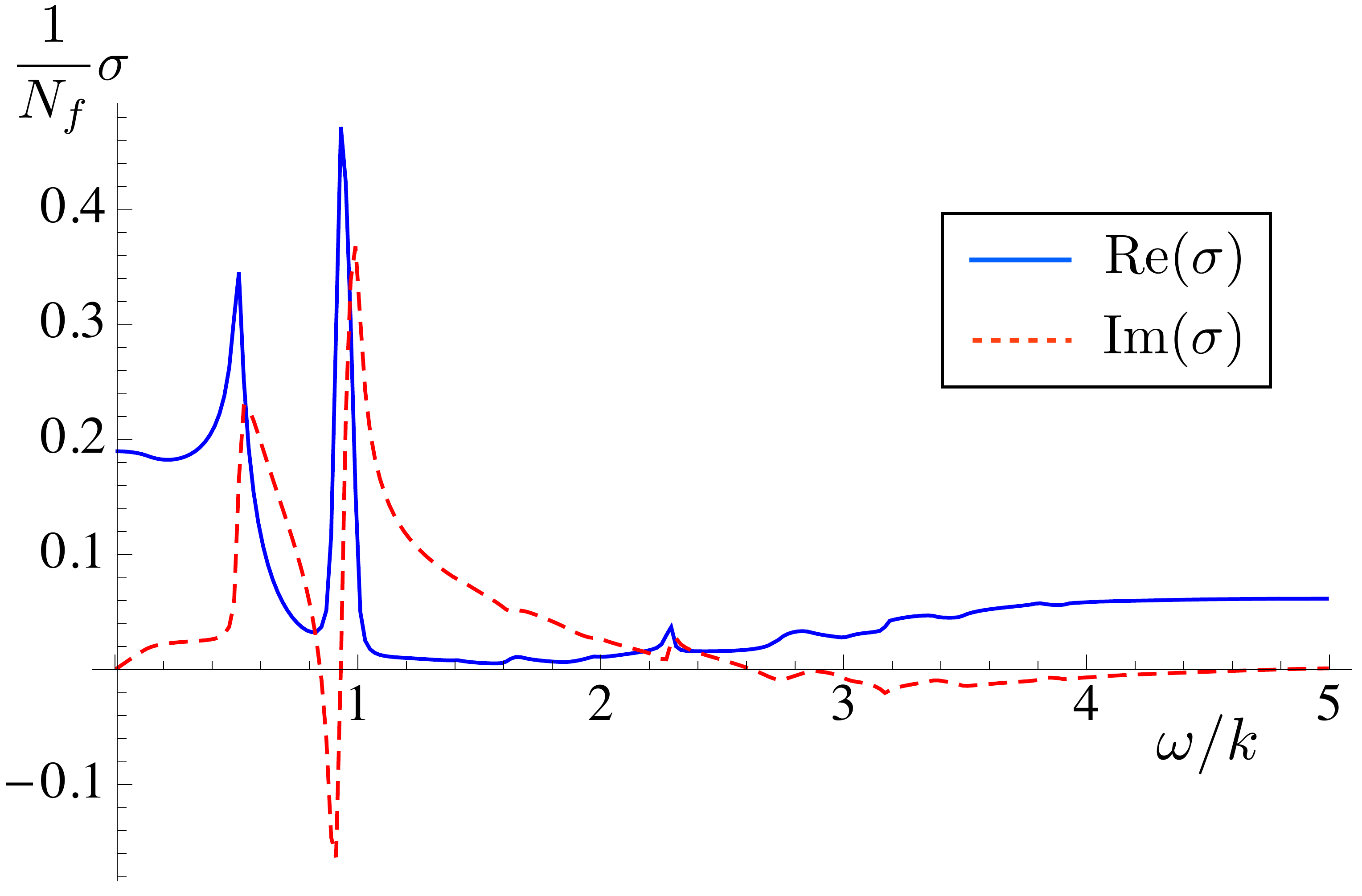}
\caption{Frequency-dependent conductivity at $V/k = 2.0$ for $N_f$ Dirac fermions in a periodic chemical potential.}
\label{fig:sigma2}
\end{center}
\end{figure}
Continuing on to large $V/k$, we consider the situation when there are 3 Dirac nodes as in Fig.~\ref{fig:diracspec}B
at $V/k=2.0$ in Fig.~\ref{fig:sigma2}. Now there is a sharp inter-band transition peak, and additional structure at $\omega \sim k$; the situation
bears similarity to the holographic result in Fig.~\ref{fig:sigmahol2}, including the dip after the peak at $\omega \sim k$. 
\begin{figure}[H]
\begin{center}
\includegraphics[width=5in]{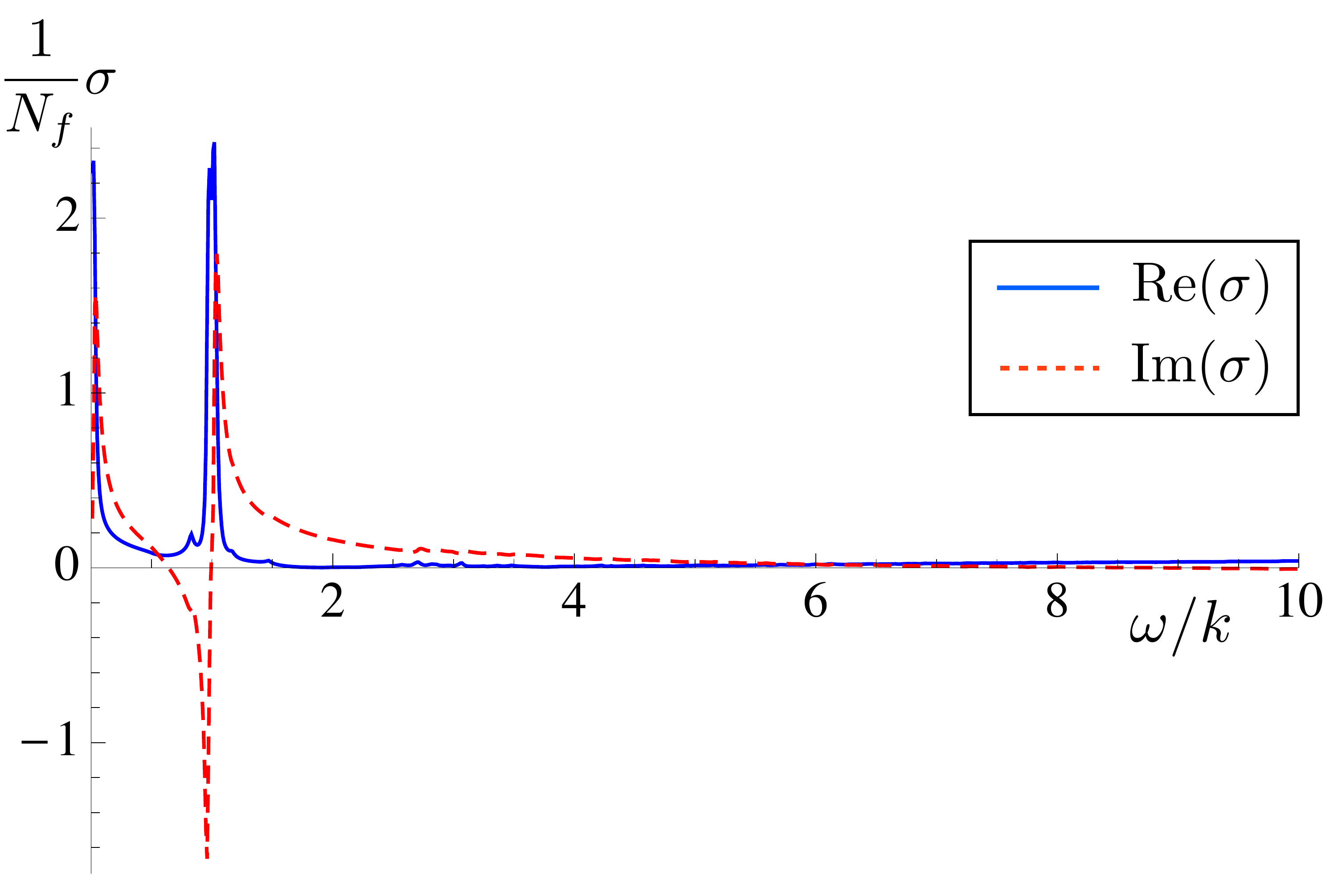}
\caption{As in Fig.~\ref{fig:sigma2}, but with $V/k = 6.0$.}
\label{fig:sigma6}
\end{center}
\end{figure}
At larger $V/k$, the peak at $\omega=k$ becomes sharper, as shown in Fig.~\ref{fig:sigma6} at $V/k=6.0$.
The reader will also notice a peak at $\omega = 0 $ in Fig.~\ref{fig:sigma6}. This arises because $V/k=6.0$ is close
to the transition point where the number of Dirac nodes jumps from 7 to 9. 

Finally, we also computed the $V/k$ dependence of $\sigma (0)$, and result showing peaks at the transition points between the IR CFTs
were shown earlier in Fig.~\ref{fig:sigmaV}. These peaks are at the points where $N_D$ jumps by 2, and the lowest energy fermions
have the dispersion in Eq.~(\ref{qcubed}).

\section{Conclusions}
\label{sec:conc}

The basic physics of 2+1 dimensional CFTs in a periodic potential is elegantly captured by Figs.~\ref{fig:squarewell} and~\ref{fig:diracpts2}.
As illustrated in Fig.~\ref{fig:squarewell}, in each local region of the potential, the CFT acquires a local Fermi surface determined by the local chemical potential. Each such region is therefore described by the theory of a Fermi surface coupled to a gauge field, as discussed recently in 
Refs.~\cite{metnem,mross,sungsik}; holographically these are `hidden' Fermi surfaces, in the language of Ref.~\cite{hss}. 
Now we go beyond the Born-Oppenheimer picture and stitch the regions together.
This by can be described by a ``folding'' of the Fermi surfaces back into the first Brillouin zone with $|q_x| < k/2$, as shown in 
Fig~\ref{fig:diracpts2}. Computations of the fermion spectrum show that new zero energy Dirac points emerge at the intersections of the folded
Fermi surfaces, represented by the filled circles in Fig~\ref{fig:diracpts2}. So if we started with a UV CFT with $N_f$ Dirac fermions,
we end up with an IR CFT with $N_D N_f$ Dirac fermions, where $N_D$ is an odd integer which increases in steps with increasing $V/k$;
for large $V/k$, $N_D$ is linearly proportional to $V/k$. These emergent Dirac points are stable under the gauge interactions,
and this stability can be regarded as a {\em remnant of the Luttinger theorem\/} for the underlying Fermi surfaces. Note that any possible ``$c$ theorem''
is badly violated because the IR CFT has many more gapless degrees of freedom; this is permitted in systems which are not
Lorentz invariant \cite{swingle}.

We computed the frequency-dependent conductivity, $\sigma (\omega)$ of CFTs in a periodic potential, both in the Dirac fermion theory
and in the Einstein-Maxwell holographic theory. For small $V/k$, the forms of $\sigma (\omega)$ are remarkably similar, as shown
in Fig.~\ref{fig:sigma0p5} and~\ref{fig:sigmahol1}. For large $V/k$, the correspondence is not as good, but the most
prominent features of $\sigma (\omega)$ do match: both the Dirac fermion theory (in Figs.~\ref{fig:sigma2} and~\ref{fig:sigma6})
and the holographic theory (in Fig.~\ref{fig:sigmahol2}) have a large peak in $\sigma (\omega)$ at $\omega \sim k$, followed by a dip
until $\omega \sim V$. A similar $\omega \sim k$ peak was also present for $\mu_0 \neq 0$, although in this case there was also a larger
Drude peak at $\omega =0 $ \cite{gary2}.

Also interesting was the dependence of the d.c. conductivity, $\sigma (0)$, on $V/k$.
This is shown in Figs.~\ref{fig:sigmaV} and~\ref{fig:sigmaholV} for the two approaches. 
While the background evolutions match, the Dirac fermion result in Fig.~\ref{fig:sigmaV} has sharp peaks at precisely
the values of $V/k$ where the integer $N_D$ jumps \cite{brey1}. This discretely increasing $N_D$, a signal of emergent zero modes
and underlying Fermi surfaces, is evidently not captured by the holographic Einstein-Maxwell theory.
It has been argued \cite{sscomp,kachru3,bao1} 
that monopole operators need to be included in the Einstein-Maxwell theory to properly quantize charge,
and to obtain signatures of the Fermi surface: so it will be interesting to add monopole operators to the present analysis and see
if they complete the correspondence between holography and field theory by leading to peaks in
 $\sigma (0)$
as function of $V/k$.

Finally, we note that the fermions dispersion spectra shown in Figs.~\ref{fig:dirac08}, \ref{fig:dirac20},  and \ref{fig:diracspec} 
apply unchanged for the case of a non-zero average chemical potential, $\mu_0 \neq 0$. The chemical potential would then not lie
precisely at the Dirac points, but away from it: so each Dirac point would have an associated Fermi surface. The Fermi surface excitations remain
coupled to the gauge field $a_\mu$, and so these Fermi surfaces are `hidden' \cite{hss} and 
they do not annihilate into Dirac points at $\mu_0 \neq 0$ . Thus the evolution of the low-energy
spectrum involves discrete jumps in the number of Fermi surfaces with increasing $V/k$. This surely has significant consequences
for the charge transport. As was the case here for $\mu_0 = 0$, we expect that the holographic method smooths out the transitions between
the changes in Fermi surface topology.

\section*{Acknowledgements}\addcontentsline{toc}{section}{Acknowledgements}

We thank H.~Fertig, S.~Hartnoll, G.~Horowitz, J.~Santos, K.~Schalm, and D.~Tong for valuable discussions.
We are especially grateful to K.~Schalm for extensive discussions in the early stages of this project.
P.C. is supported by a Pappalardo Fellowship in Physics at MIT.   A.L.  is supported by the Purcell Fellowship at Harvard University.  This research was supported by the US National Science Foundation
under Grant DMR-1103860 
and by a grant from the John Templeton Foundation. This research was also supported in part by Perimeter Institute for
Theoretical Physics; research at Perimeter Institute is supported by the
Government of Canada through Industry Canada and by the Province of
Ontario through the Ministry of Research and Innovation. S.S. acknowledges 
the hospitality of the Simons Symposium on ÔQuantum Entanglement: From Quantum Matter to String TheoryÕ.

\begin{appendix}
\titleformat{\section}
  {\gdef\sectionlabel{}
   \Large\bfseries\scshape}
  {\gdef\sectionlabel{\thesection. }}{0pt}
  {\begin{tikzpicture}[remember picture,overlay]
	\draw (-0.2, 0) node[right] {\textsf{Appendix \sectionlabel#1}};
	\draw[thick] (0, -0.4) -- (\textwidth, -0.4);
       \end{tikzpicture}
  }
\titlespacing*{\section}{0pt}{15pt}{20pt}

\section{Results from Perturbative Holography}\label{appa}
This appendix collects the cumbersome calculations required to derive the first order perturbation to the conductivity and to the geometry in the limit $V\ll k$.
\subsection{Geometry}
We begin with the perturbation theory for the metric, which we find by computing the solution to the linearized graviton equation of motion.   For simplicity, we do this calculation in Euclidean time, and analytically continue at the end, as we will do for much of this appendix.   In the transverse traceless gauge ($\nabla^\mu h_{\mu\nu}=0$), higher spin equations of motion take a particularly elegant form \cite{xiyin}, which for spin 2 reads (using our conventions of $L=1$, etc.): \begin{equation}
\nabla_\rho \nabla^\rho h_{\mu\nu} + 2 h_{\mu\nu} = - T^{(\mathrm{EM})}_{\mu\nu},
\end{equation}where now we treat $T^{(\mathrm{EM})}_{\mu\nu}$ as the source term for the gravitons.   The traceless condition is allowed because electromagnetic fields in 4 dimensions have a traceless stress tensor.     We also note that the transverse gauge implies that \begin{equation}
\left(\partial_z - \frac{2}{z}\right)h_{z\mu} + \partial_i h_{i\mu} = 0.   \label{transv}
\end{equation}

Let us first study the stress tensor.    In Euclidean time, the photon is $A = \mathrm{i}V \mathrm{e}^{-kz}\cos(kx)\mathrm{d}t$.   This leads to \begin{equation}
F = \mathrm{d}A = \mathrm{i}V k \mathrm{e}^{-kz}\left[\cos(kx)\mathrm{d}x\wedge\mathrm{d}t - \sin(kx)\mathrm{d}z\wedge\mathrm{d}t\right]
\end{equation}and correspondingly, the only non-zero components of the stress tensor are\begin{subequations}\begin{align}
\frac{1}{z^2}T^{(\mathrm{EM})}_{zz}  &= -\frac{1}{z^2}T^{(\mathrm{EM})}_{xx}= F_{zt}^2- \frac{1}{2}F_{zt}^2 - \frac{1}{2}F_{xt}^2 = \frac{F_{zt}^2 - F_{xt}^2}{2} = \frac{1}{2}V^2k^2\cos(2kx)\mathrm{e}^{-2kz}. \\
\frac{1}{z^2}T^{(\mathrm{EM})}_{tt}  &= - \frac{1}{z^2}T^{(\mathrm{EM})}_{yy}  =  \frac{F_{xt}^2 + F_{zt}^2}{2} = -\frac{1}{2}V^2k^2 \mathrm{e}^{-2kz} \\
\frac{1}{z^2}T^{(\mathrm{EM})}_{zx} &= F_{zt}F_{xt} = -\frac{1}{2}V^2k^2\sin(2kx)\mathrm{e}^{-2kz}
\end{align}\end{subequations}
Then, using that \begin{subequations}\label{graveqs}\begin{align}
\nabla_\rho\nabla^\rho h_{zz} + 2h_{zz} &= \left(z^2 \partial_z^2 - 2z\partial_z  + z^2\partial_k^2\right) h_{zz} \\
\nabla_\rho\nabla^\rho h_{zi} + 2h_{zi} &= \left(z^2 \partial_z^2 - 2+ z^2\partial_k^2\right) h_{zi} - 2r\partial_i h_{zz} \\
\nabla_\rho\nabla^\rho h_{ij} + 2h_{ij} &= \left(z^2 \partial_z^2 + 2z \partial_z  - 2+ z^2\partial_k^2\right) h_{ij} - 2z\partial_i h_{jz}- 2z\partial_j h_{iz} + 2\delta_{ij}h_{zz} 
\end{align}\end{subequations}along with the fact that the non-zero components of the stress tensor are $rr, rx, xx, tt, yy$, we conclude that $h_{tx}=h_{ty}=h_{xy}=h_{zt}= h_{zy}=0$.     Note that the boundary conditions on the non-zero perturbations are that they vanish on the conformal boundary of AdS ($z=0,\infty$).

Now, let us turn to the solutions of these equations of motion.   By studying the stress tensor, one concludes that the solutions to the equations of motion with the proper boundary conditions will be of the form \begin{subequations}\begin{align}
h_{zz}(z,x) &= H_{zz}(z)\cos(2kx), \\
h_{zx}(z,x) &= H_{zx}(z)\sin(2kx), \\
h_{xx}(z,x) &= H_{xx}(z)\cos(2kx), \\
h_{yy}(z,x) &=  G_{yy}(z) + H_{yy}(z)\cos(2kx), \\
h_{tt}(z,x) &= -G_{yy}(z)+H_{yy}(z)\cos(2kx).
\end{align}\end{subequations}Furthermore, since these are linear equations, $G_{yy}$ decouples from the rest.   We note from Eq. (\ref{graveqs}) that $H_{zz}$ will act as a source for $H_{zx}$ and $H_{yy}$; we will find $H_{xx}$ by a trick.     

Let us begin with $G_{yy}(z)$, which obeys the equation of motion \begin{equation}
-\frac{V^2k^2}{2}\mathrm{e}^{-2kz} = \frac{\mathrm{d}^2 G_{yy}}{\mathrm{d}z^2} + \frac{2}{z}\frac{\mathrm{d}G_{yy}}{\mathrm{d}z} - \frac{2G_{yy}}{z^2},
\end{equation}which has exact solution \begin{equation}
G_{yy}(z) = V^2 \frac{1-\mathrm{e}^{-2kz}\left(1+2kz+2k^2z^2\right)}{16k^2z^2}.   \label{exactgyy}
\end{equation}Eq. (\ref{exactgyy}) has asymptotic behavior in the UV \begin{equation}
G_{yy}(r) \approx \frac{V^2 kz}{12},   \label{4stress1}
\end{equation}which we will find to be a useful fact later.

The next step is to compute $H_{zz}(z)$, which obeys an independent equation: \begin{equation}
-\frac{V^2k^2}{2}\mathrm{e}^{-2kz} = \frac{\mathrm{d}^2H_{zz}}{\mathrm{d}z^2 } - \frac{2}{z}\frac{\mathrm{d}H_{zz}}{\mathrm{d}z} - 4k^2 H_{zz}.
\end{equation}The boundary conditions are that $H_{zz}(0)=H_{zz}(\infty)=0$.   The solution to this differential equation may be found exactly: \begin{align}
H_{zz}(z) &=  \frac{V^2}{16}\left[\mathrm{e}^{-2kz}\left(2+(1+2kz)\log(kz)\right) + (2kz-1)\mathrm{e}^{2kz}\mathrm{Ei}(-4kz)\right]\notag \\
&\;\;\;\;\;\; + \frac{V^2\left(\log 4 + \gamma - 2\right)}{16}(1+2kz) \mathrm{e}^{-2kz}.
\end{align}where $\gamma\approx 0.577$ is the Euler-Mascheroni constant and \begin{equation}
\mathrm{Ei}(x) \equiv -\int\limits_{-x}^\infty \mathrm{d}t \; \frac{\mathrm{e}^{-t}}{t}.
\end{equation}Note that $\mathrm{Ei}(-4kz)\sim \mathrm{e}^{-4kz}$ up to polynomial size corrections, at large $z$, and so we have $H_{zz} \sim \mathrm{e}^{-2kz}$ in the IR as expected.   In the UV, we have \begin{equation}
H_{zz}(z) \approx \frac{V^2 k^2 z^2}{4}.
\end{equation}

Now let us turn to $H_{zx}$ and $H_{xx}$.   We can actually compute these very ``efficiently" using the transverse gauge condition Eq. (\ref{transv}).   We will not bother writing down the exact solution, but we will extract the leading order behavior at small $r$, which requires the next order in the asymptotic expansion of $H_{zz}(z)$, which is of order $z^3\log (kz)$.   One can show that\begin{equation}
H_{zx}(z) \approx \frac{V^2 k^2 }{6} z^2 \log (kz).
\end{equation}A second application of Eq. (\ref{transv}) immediately leads to asymptotic behavior \begin{equation}
H_{xx}(z) \approx \frac{V^2 kz}{12}. \label{4stress2}
\end{equation}The traceless condition then leads to \begin{equation}
H_{yy}(z) \approx - \frac{V^2 kz}{24}.  \label{4stress3}
\end{equation}This concludes the derivation of the first order perturbation to the metric.

\subsection{Geometry in Numerics Gauge}
One of the major checks of the numerics is to compare the geometry to this perturbative answer, and more importantly the expectation value of the stress tensor, which is gauge invariant.    To show how this can be done, let us describe explicitly a sequence of gauge transformations which will take us from the equations of the previous subsection to a metric of the form Eq. (\ref{numericform}).

Since we are looking for a perturbative geometry around AdS, let us expand the above metric to lowest order about AdS, where $B_0=F_0=0$,  $A_0=\Sigma_0^2$ and $\Sigma_0 = r^{-1}$.    If we let $A=\Sigma^2+A_1$,  $B=r^2B_1$, $F=F_1$ and $\Sigma=\Sigma_0+\Sigma_1$, we find \begin{equation}
\mathrm{d}s^2 =-\left(\frac{1}{z^2} + \frac{2\Sigma_1}{z}+A_1\right)\mathrm{d}t^2 - \frac{2\mathrm{d}z\mathrm{d}t}{z^2} + 2F_1\mathrm{d}x\mathrm{d}t + \left(\frac{1}{z^2}+\frac{2\Sigma_1}{z}+B_1\right)\mathrm{d}x^2  + \left(\frac{1}{z^2}+\frac{2\Sigma_1}{z}-B_1\right)\mathrm{d}y^2 
\end{equation}
The first thing we can do is make the simple coordinate change $t\rightarrow \mathrm{i}t$ (back to real time) and $t\rightarrow t+z$ to put the metric in the form: \begin{align}
\mathrm{d}s^2 &= \frac{((H_{zz}-H_{yy})\cos(2kx)+G_{yy})z^2\mathrm{d}z^2 - (1-z^2G_{yy}+z^2H_{yy}\cos(2kx))(\mathrm{d}t^2 + 2\mathrm{d}z\mathrm{d}t)}{z^2} \notag \\
&+ \frac{2H_{zx}\sin(2kx)z^2\mathrm{d}x\mathrm{d}z + (1+G_{yy}z^2+H_{yy}z^2\cos(2kx))\mathrm{d}y^2 +(1+H_{xx}z^2\cos(2kx))\mathrm{d}x^2}{z^2}
\end{align}
Up to $\mathrm{O}(V^2)$ terms, this is of the form of the numerical metric.     So our remaining task will be to find the $\mathrm{O}(V^2)$ diffeomorphism to get the rest right -- we will drop all higher terms.

Let's begin by shifting $t$ again to remove the $\mathrm{d}r^2$ term.   Let a $(-1)$ superscript represent integration of a function multiplied by $r^2$:  e.g., $\mathrm{d}H^{(-1)}/\mathrm{d}z = z^2H$:   then we find that if we choose \begin{equation}
t\rightarrow t + \frac{(H_{zz}^{(-1)}-H_{yy}^{(-1)})\cos(2kx) + G_{yy}^{(-1)}}{2}
\end{equation}we can set $g_{zz}=0$ and shift \begin{equation}
z^2 g_{zt} = 1+\frac{(H_{rr}+H_{yy})\cos(2kx)-G_{yy}}{2}z^2
\end{equation}\begin{equation}
z^2 g_{tx} = k(H_{zz}^{(-1)}-H_{yy}^{(-1)})\sin(2kx).
\end{equation}

Next we remove $g_{xz}$ by shifting $x$ (which must remain periodic with period $2\pi/k$)  by a periodic function: \begin{equation}
x \rightarrow x - H_{zx}^{(-1)}\sin(2kx)
\end{equation}which removes $g_{xr}$ and changes \begin{equation}
z^2 g_{xx} = 1+(2kH_{zx}^{(-1)}+z^2H_{xx})\cos(2kx)
\end{equation}

The final step is to change $g_{zt}=1$.   To do this we perform a coordinate shift on $z$ so that \begin{equation}
\mathrm{d}\frac{1}{z^\prime} =z^2 g_{zt}\mathrm{d}\frac{1}{z}
\end{equation}Since to lowest order we can set $z=z^\prime$ in $z^2g_{zt}$, we find that \begin{equation}
\frac{1}{z}\rightarrow \int\limits_0^z  \frac{\mathrm{d}z^\prime}{z^{\prime 2} z^2g_{zt}} \approx \int\limits_z^\infty \frac{\mathrm{d}z^\prime}{z^{\prime 2}} \left[1-z^2\frac{(H_{zz}+H_{yy})\cos(2kx)-G_{yy}}{2}\right] = \frac{1}{z} + \frac{(H_{rr}^{[-1]} + H_{yy}^{[-1]})\cos(2kx) - G_{yy}^{[-1]}}{2}
\end{equation}where $\mathrm{d}G^{[-1]}/\mathrm{d}z =G$ (note the sign on the last term to ensure that the derivative factor works out right).
This sets $g_{zt}=1$, and it will also multiply all \emph{other} metric components by the factor corresponding to the shift in $1/z^2$: \begin{equation}
\frac{1}{z^2} \rightarrow \frac{1}{z^2} -  \frac{(H_{zz}^{[-1]} + H_{yy}^{[-1]})\cos(2kx) - G_{yy}^{[-1]}}{z}
\end{equation}

Straightforward manipulations then give: \begin{subequations}\begin{align}
B_1 &= \frac{g_{xx}-g_{yy}}{2} = \frac{1}{2}\left[ \left(\frac{1}{z^2} -  \frac{(H_{zz}^{[-1]} + H_{yy}^{[-1]})\cos(2kx) - G_{yy}^{[-1]}}{z}+\left(H_{xx}+\frac{2kH_{rx}^{(-1)}}{z^2}\right)\cos(2kx)\right)\right. \notag \\
&\left. - \left(\frac{1}{z^2} -  \frac{(H_{rr}^{[-1]} + H_{yy}^{[-1]})\cos(2kx) - G_{yy}^{[-1]}}{z}+G_{yy}+H_{tt}\cos(2kx)\right)\right] \notag \\
&=\frac{-G_{yy}+(H_{xx} + 2z^{-2}kH^{(-1)}_{zx}-H_{yy})\cos(2kx)}{2} \\
F_1 &= g_{tx} = \frac{k(H_{zz}^{(-1)} - H_{yy}^{(-1)})\sin(2kx)}{z^2} \\
\Sigma_1 &= \frac{z(g_{xx}+g_{yy}-2z^{-2})}{4} = \frac{z}{4}\left[\left(\frac{1}{z^2} -  \frac{(H_{zz}^{[-1]} + H_{yy}^{[-1]})\cos(2kx) -G_{yy}^{[-1]}}{z}+G_{yy}+H_{yy}\cos(2kx)\right) \right. \notag \\
&\left. +\left( \frac{1}{z^2}-  \frac{(H_{zz}^{[-1]} + H_{yy}^{[-1]})\cos(2kx) - G_{yy}^{[-1]}}{z}+\left(H_{xx}+\frac{2kH_{zx}^{(-1)}}{z^2}\right)\cos(2kx)\right) - \frac{2}{z^2}\right] \notag \\  &= -\frac{(H_{zz}^{[-1]} + H_{yy}^{[-1]})\cos(2kx) - G_{yy}^{[-1]}}{2} +\frac{z}{4}\left[\left(H_{xx}+H_{yy} + \frac{2k}{z^2}H^{(-1)}_{zx}\right)\cos(2kx)+G_{yy}\right] \\
A_1 &= -g_{tt} - \frac{1}{z^2}-\frac{2\Sigma_1}{z} = \frac{1}{z^2}-\frac{(H_{zz}^{[-1]} + H_{yy}^{[-1]})\cos(2kx) - G_{yy}^{[-1]}}{z} - G_{yy} + H_{yy}\cos(2kx) \notag \\
& - \frac{1}{z^2} +\frac{(H_{zz}^{[-1]} + H_{yy}^{[-1]})\cos(2kx) - G_{yy}^{[-1]}}{z} -\frac{1}{2}\left[\left(H_{xx}+H_{yy} + \frac{2k}{z^2}H^{(-1)}_{zx}\right)\cos(2kx)+G_{yy}\right] \notag \\
&= -\frac{1}{2}\left[\left(H_{xx}-H_{yy} + \frac{2k}{z^2}H^{(-1)}_{zx}\right)\cos(2kx)+3G_{yy}\right]
\end{align}\end{subequations}

\subsection{Conductivity}
In this subsection we compute $\sigma_{\mathrm{el}}(\omega)$ and $\sigma_{\mathrm{inel}}(\omega)$.   As before, it will be easier to perform this calculation in Euclidean time, when all integrals are well-behaved and convergent.   At the end of the calculation, we will do an analytic continuation $\omega\rightarrow -\mathrm{i}\omega$ to extract the real time conductivity.

The strategy will be to compute the three Witten diagrams of Figure \ref{witten}.   The photon and graviton propagators and vertices for an $\mathrm{AdS}_4$ background can be found in \cite{chowdhury}, in real time.    Let us begin by reviewing them in Euclidean time.   For the probe $A_x$ field, normalized to 1 on the boundary so we are computing the conductivity, we have \begin{equation}
\langle A_x(\omega,z)A_x(-\omega,0)\rangle =  \mathrm{e}^{-|\omega|z}.
\end{equation}\begin{equation}
\langle A_t(k,z)A_t(-k,0)\rangle = \mathrm{i}\frac{V}{2} \mathrm{e}^{-|k|z}  
\end{equation}and for the graviton we have, in the axial gauge where $h_{z\mu}=0$: \begin{equation}
\langle h_{ij}(q_x,q_t,z_1)h_{kl}(-q_x,-q_t,z_2)\rangle = -\frac{1}{8}\int\limits_0^\infty \frac{\mathrm{d}x}{\sqrt{z_1z_2}} \frac{\mathrm{J}_{3/2}(\sqrt{x}z_1)\mathrm{J}_{3/2}(\sqrt{x}z_2)}{x+q_x^2+q_t^2} \left(\mathcal{T}_{ik}\mathcal{T}_{jl}+\mathcal{T}_{il}\mathcal{T}_{jk}-\mathcal{T}_{ij}\mathcal{T}_{kl}\right)
\end{equation}where $\mathrm{J}_{3/2}$ is a Bessel function, and \begin{equation}
\mathcal{T}_{ij} = \delta_{ij}+\frac{q_iq_j}{x}.
\end{equation}
The interaction vertex comes from the terms in the action \begin{equation}
\frac{1}{2}\left[F^{\mu\rho}{F^\nu}_\rho h_{\mu\nu} - \frac{1}{4}h^\mu_\mu F_{\rho\sigma}F^{\rho\sigma}\right].
\end{equation}The factor of 1/2 comes from the fact that the usual definition of the stress tensor is $2g^{-1/2}\delta S/\delta g_{\mu\nu}$.   Using the background metric of AdS given by Eq. (\ref{g0}), we can simplify the interaction vertex to (now there are no metric factors in summation convention:  simply sum): \begin{equation}
\frac{z^2}{2} \left[F_{\mu\rho}F_{\nu\rho}h_{\mu\nu} - \frac{1}{4}h_{\mu\mu}F_{\rho\sigma}F_{\rho\sigma}\right]
\end{equation}

 Let us begin with the elastic channel: $A_tA_th\rightarrow A_xA_xh$ (here $h$ represents a generic graviton).   In this case, the graviton carries no internal momentum, so $q_t=q_x=0$ and the vertex will be a bit simpler.   What are the possible internal modes of the graviton?   At the $A_tA_th$ vertex, we have $F_{zt} = -|k|A_t$ and $F_{xt} = \mathrm{i}kA_t$.   Now since we have $A_t(k)A_t(-k)$, the only surviving terms are: \begin{equation}
\frac{z^2}{2} \left[F_{tz}F_{tz}h_{tt} + F_{tx}F_{tx}h_{tt} + F_{tx}F_{tx}h_{xx} - \frac{1}{2}(h_{tt}+h_{xx}+h_{yy})(F_{xt}^2 + F_{zt}^2)\right]
\end{equation}We need to take into account a symmetry factor of 2 because there are 2 photons in the same direction, so we will for simplicity add that to the vertex factor now.   We find \begin{equation}
z^2k^2 A_t A_t\left[2h_{tt} + h_{xx} - \frac{2}{2}(h_{tt}+h_{xx}+h_{yy})\right] = z^2k^2 A_tA_t(h_{tt}-h_{yy}).
\end{equation}It should be clear that this calculation is exactly the same for $A_xA_x h$, with an appropriate exchange of the $t$ and $x$ labels, as well as the $\omega$ and $k$ factors: \begin{equation}
z^2\omega^2 A_xA_x (h_{xx}-h_{yy}).
\end{equation}

Now, let us look at the total $\mathcal{T}_{ij}$ dependent factor -- let's call it $\alpha$.   We will also include in this polarization factor the vertex coefficients.   In this case, it is quite simple: \begin{equation}
\alpha = \alpha_{tt,xx}+\alpha_{yy,xx}  + \alpha_{tt,yy}+\alpha_{yy,yy}
\end{equation}where $\alpha_{tt,xx}$ corresponds to $A_t A_t h_{tt} \rightarrow A_xA_xh_{xx}$, e.g. \begin{subequations}\begin{align}
\alpha_{tt,xx} &= \left(2\mathcal{T}_{xt}^2 - \mathcal{T}_{tt}\mathcal{T}_{xx}\right) \times \omega^2\times k^2 = -k^2\omega^2 \\
\alpha_{yy,xx} &= (-1)\times -\omega^2 \times k^2 = k^2\omega^2 \\
\alpha_{yy,yy} &= (2-1)\times -\omega^2 \times -k^2 = \omega^2k^2 \\
\alpha_{tt,yy} &= (-1)\times -\omega^2\times k^2 = \omega^2k^2.
\end{align}\end{subequations}So we find the polarization/vertex reduces to \begin{equation}
\alpha = 2\omega^2k^2.
\end{equation}So now we simply attach the boundary-bulk propagators, and account for a remaining symmetry factor of 2 because either $A_t$ can be treated as incoming.   A remaining factor of 2 is needed due to the normalization of the diagram with four external legs.  Thus we find our first correction to the conductivity: \begin{equation}
\omega \sigma_{\mathrm{el}} = -\frac{1}{8} \int  \mathrm{d}x \mathrm{d}z_1 \mathrm{d}z_2  (z_1z_2)^{3/2} (\mathrm{i}^2V^2 \mathrm{e}^{-2\omega z_1}\mathrm{e}^{-2kz_2})(2\omega^2k^2) \frac{\mathrm{J}_{3/2}(\sqrt{x}z_1)\mathrm{J}_{3/2}(\sqrt{x}z_2)}{x}
\end{equation}
Now, we use the integral \begin{equation}
\int\limits_0^\infty \mathrm{d}z\; z^{3/2} \mathrm{J}_{3/2}(az)\mathrm{e}^{-bz} = \sqrt{\frac{8}{\pi}} \frac{a^{3/2}}{(a^2+b^2)^2}
\end{equation}and find \begin{equation}
\sigma_{\mathrm{el}} = \frac{2}{\pi}V^2\omega k^2\int\limits_0^\infty \mathrm{d}x \frac{x^{3/2}}{x(x+4\omega^2)^2(x+4k^2)^2} = \frac{V^2 k}{32(k+\omega)^3}.
\end{equation}Analytic continuation is straightforward and gives the first half of Eq. (\ref{2sigma}).

Next we compute $\sigma_{\mathrm{inel}}$, which corresponds to $A_xA_t h \rightarrow A_xA_t h$.   This one is more subtle because of the graviton polarization factor.   To begin, let us analyze the vertex: \begin{equation}
\frac{r^2}{2} \left[2h_{xt}F_{xz}F_{tz} + (h_{xx}+h_{tt})F_{xt}^2 - \frac{1}{2}(h_{xx}+h_{tt}+h_{yy})F_{xt}^2 \right]
\end{equation}Now since $F_{xt} = \mathrm{i}q_x A_t + \mathrm{i}q_t A_x$, we extract $-2q_xq_t A_x A_t$ from $F_{xt}^2$ (remember we need an $A_x$ and an $A_t$).  This turns the vertex into \begin{equation}
\frac{z^2}{2} \left[2h_{xt} \omega k A_x A_t + (h_{yy}-h_{xx}-h_{tt}) q_xq_t A_xA_t\right].
\end{equation}Note that $q_x = \pm k$ and $q_t = \pm \omega$, but it will be very convenient to not plug in yet.   

Now let us look at the polarization factor $\alpha$.   There are a lot of terms to take into account, but we can eliminate some early.   Suppose we take a term of the form $\omega k q_x q_t$.   We have two diagrams to sum over, one where the photons on one side have $+\omega$ and $+k$, and one where the photons have $+\omega$ and $-k$.   (Flipping the sign on both $\omega$ and $k$ just means flipping the diagram, and this does not count as a separate diagram).   So anything odd in $q_x$ (or $q_t$) vanishes under the sum of these two diagrams.   

This means we only have to consider $h_{xt}\rightarrow h_{xt}$ and then the mixing of the diagonal $h$s.    We have \begin{subequations}\begin{align}
\alpha_{xt,xt} &= 2\left(\mathcal{T}_{xx}\mathcal{T}_{tt}+(1-1)\mathcal{T}_{xt}^2\right)\times \omega k \times \omega k = 2\omega^2k^2 \left(1+\frac{k^2}{x}\right)\left(1+\frac{\omega^2}{x}\right) \\
\alpha_{xx,xx} &= \frac{1}{2}\mathcal{T}_{xx}^2 \times (-q_xq_t)^2 = \frac{1}{2}\omega^2k^2\left(1+\frac{k^2}{x}\right)^2 \\
\alpha_{tt,tt} &= \frac{1}{2}\mathcal{T}_{tt}^2 \times (-q_xq_t)^2 = \frac{1}{2}\omega^2k^2\left(1+\frac{\omega^2}{x}\right)^2 \\
\alpha_{xx,tt} &=\left(2\mathcal{T}_{xt}^2 - \mathcal{T}_{tt}\mathcal{T}_{xx}\right)(-q_xq_t)^2 = \omega^2k^2 \left(\frac{\omega^2k^2}{x^2} - 1 - \frac{\omega^2+k^2}{x}\right) \\
\alpha_{yy,yy} &= \frac{1}{2}\mathcal{T}_{yy}^2 \times (q_xq_t)^2 = \frac{1}{2}\omega^2k^2 \\
\alpha_{yy,xx} &= (-\mathcal{T}_{xx}\mathcal{T}_{yy}) \times q_xq_t \times -q_xq_t = \omega^2k^2\left(1+\frac{k^2}{x}\right) \\
\alpha_{yy,tt} &= (-\mathcal{T}_{tt}\mathcal{T}_{yy}) \times q_xq_t \times -q_xq_t = \omega^2k^2\left(1+\frac{\omega^2}{x}\right) 
\end{align}\end{subequations}Note when the gravitons have 2 different components, there is an extra factor of 2 by symmetry (which side has which component).   We get \begin{equation}
\alpha = \frac{1}{2} \left[9 + 6\frac{\omega^2+k^2}{x} +   \frac{\left(\omega^2+k^2\right)^2}{x^2} + 4\frac{\omega^2k^2}{x^2}\right]\omega^2k^2   \label{alpha2}
\end{equation}
We find the diagram, accounting for symmetry factors similarly to before: \begin{equation}
\omega \sigma_{\mathrm{inel}} = \frac{V^2 \omega^2k^2}{4} \int\limits_0^\infty \mathrm{d}x \mathrm{d}z_1 \mathrm{d}z_2 \; (z_1z_2)^{3/2} \mathrm{e}^{-(k+\omega)(z_1+z_2)} \frac{\mathrm{J}_{3/2}(\sqrt{x}z_1) \mathrm{J}_{3/2}(\sqrt{x}z_2)}{x+\omega^2+k^2} \alpha(x),
\end{equation} which leads to\begin{equation}
\sigma_{\mathrm{inel}} = \frac{2}{\pi \omega} V^2 k^2 \int\limits_0^\infty \mathrm{d}x \frac{x^{3/2}}{(x+\omega^2+k^2)(x+(\omega+k)^2)^4} \alpha(x).   \label{eucsigmainel}
\end{equation}This integral can be done with the aid of symbolic manipulators (Mathematica) and the result, appropriately analytically continued, is the second half of Eq. (\ref{2sigma}).   Note that the rightmost term in Eq. (\ref{alpha2}) is responsible for the singularity in the real-time $\sigma_{\mathrm{inel}}$.

\section{Dirac fermions in a periodic potential}
\label{app:dirac}

We describe the numerical computation of the conductivity of a single Dirac fermion in a periodic potential. 

We write the ``reciprocal lattice vector'' of the periodic potential as ${\bm K} = (k, 0)$.
We introduce 2 canonical fermions $c_A (\bq + \ell {\bm K})$ and $c_B (\bq + \ell {\bm K})$ where $\ell$ is an integer,
and $\bq$ is restricted to the ``first Brillouin zone'', $|k_x| < k/2$. We discretized the momenta $\bq = (k/N) (n_x, n_y)$
where $n_x$, $n_y$, and $N$ are integers, with $|n_x| < N/2$, $|n_y| < N_y$, and $|\ell| < L$. The continuum answers are
obtained in the limit where $N$, $N_y$, and $L$ all become very large.

The Hamiltonian of the Dirac fermions in a periodic potential now takes the form
\begin{eqnarray}
H &=& \sum_{\ell=0} {\sum_{\bq}} \Biggl[ \left( T(\bq + \ell {\bm K}) 
c_{A}^\dagger (\bq + \ell {\bm K}) c_{B}^{\vphantom \dagger} (\bq + \ell {\bm K}) + \mbox{c.c.} \right) \nn
&~& + \frac{V}{2}  \left( c_{A}^\dagger (\bq + (\ell + 1) {\bm K}) c_{A}^{\vphantom \dagger} (\bq + \ell {\bm K}) + 
 c_{B}^\dagger (\bq + (\ell + 1){\bm K}) c_{B}^{\vphantom \dagger} (\bq+ \ell {\bm K}) \right) + \mbox{c.c.} \Biggr]
\end{eqnarray}
The bare kinetic energy term is
\beq
T (\bq) = q_x - \mathrm{i} q_y
\eeq
obtained from the Pauli matrices acting as the Dirac matrices.
We diagonalize this Hamiltonian by the unitary
\beq
c_{\sigma} (\bq + \ell {\bm K}) = \sum_{n} U_{\sigma n} (\bq + \ell {\bm K}) \gamma_n (\bq) \label{cgamma}
\eeq
where $\sigma = A, B$, $U_{\sigma n} (\bq) $ is a unitary matrix,  and $\varepsilon_n (\bq) $ are the corresponding eigenvalues.

To obtain the conductivity, we need the current operator
\beq
{\bm j} = \sum_{\ell} {\sum_\bq} c_{A}^\dagger (\bq + \ell{\bm K}) {\bm J} (\bq+ \ell {\bm K}) c_{B}^{\vphantom\dagger} (\bq+ \ell {\bm K}) + \mbox{c.c.}
\eeq
where
\beq
{\bm J} = (1, -\mathrm{i})
\eeq
is obtained from the Dirac matrices.
We introduce the matrix element of the current operator 
\beq
\mathcal{J}_{i,nm} (\bq) =  \sum_{\ell} \Bigg[
U^\ast_{A n} (\bq + \ell {\bm K} ) J_{i } (\bq+ \ell {\bm K} ) U_{B m} (\bq+ \ell {\bm K} )  + U^\ast_{B n} (\bq+ \ell {\bm K} ) J_i^\ast (\bq+ \ell {\bm K} ) U_{A m} (\bq+ \ell {\bm K} ) \Biggr]
\eeq
and the final expression for the conductivity is 
\begin{eqnarray}
&~&\sigma_{ii} (\omega) = \lim_{\eta \rightarrow 0} \lim_{N, N_y, L \rightarrow \infty} \frac{i k}{4 \pi^2 N^2 \omega } 
\left[ \Lambda (\omega) - \mbox{Re}\left( \Lambda (0) \right) \right] \nn
&~&\mbox{with } \Lambda(\omega) = {\sum_\bq} \sum_{n,m} 
|\mathcal{J}_{i,nm} (\bq) |^2 
  \frac{\left( f( \varepsilon_n (\bq)) - f( \varepsilon_m (\bq)) \right)}{\omega - \varepsilon_m (\bq) + \varepsilon_n (\bq) + i \eta} ,
  \label{sigmasub}
\end{eqnarray}
where $f(\varepsilon)$ is the Fermi function. Here $\eta$ is small energy broadening parameter, and the order of limits above is important.
Our use of a sharp momentum space cutoff, $N_y$, requires the subtraction scheme in Eq.~(\ref{sigmasub}) to eliminate cutoff dependence. With a gauge-invariant cutoff we would have $\Lambda (0)=0$, but instead we obtain a real part which diverges linearly with
the cutoff momentum. This cutoff dependence only influences $\mbox{Im} (\sigma)$, and cutoff-independent results are obtained
after the subtraction.

\subsection{Emergence of Dirac zeros}
\label{app:zeros}

This appendix will review the arguments \cite{barbier1,wang1} for the appearance of the 
additional Dirac zeros for the case of the periodic rectangular-wave potential in 
Fig.~\ref{fig:squarewell}. 

It is useful to first tabulate some properties of the Dirac equation for the case of a piecewise-constant potential.
In any given constant potential region a solution with energy $E = (k_x^2 + k_y^2)^{1/2}$ will be a superposition of plane
waves with wavevectors $(k_x, k_y)$ and $(-k_x, k_y)$ (the wavevector $k_y$ is conserved because the potential is independent of $y$).
By working with the most general linear combination of such waves, we can relate the two components of the wavefunction
$(\psi_A (x), \psi_B (x))$ between two $x$ locations by a linear transfer matrix of the form
\beq
\psi(x_1) = \mathcal{T}_+ (x_1 - x_2 ; k_x, k_y) \psi (x_2)
\eeq
It is easy to compute that the explicit form of this transfer matrix is 
\beq
\mathcal{T}_+ (x, k_x, k_y) = \left(
\begin{array}{cc}
\cos(k_x x) + (k_y/k_x) \sin(k_x x) 
& \mathrm{i} (1 + (k_y/k_x)^2)^{1/2} \sin (k_x x) \\
\mathrm{i} (1 + (k_y/k_x)^2)^{1/2} \sin (k_x x)
& \cos(k_x x) - (k_y/k_x) \sin(k_x x) 
\end{array} \right)
\eeq
Similarly, for the solution with energy $E = - (k_x^2 + k_y^2)^{1/2}$ the transfer matrix is
\beq
\mathcal{T}_- (x, k_x, k_y) = \left(
\begin{array}{cc}
\cos(k_x x) + (k_y/k_x) \sin(k_x x) 
& -\mathrm{i} (1 + (k_y/k_x)^2)^{1/2} \sin (k_x x) \\
-\mathrm{i} (1 + (k_y/k_x)^2)^{1/2} \sin (k_x x)
& \cos(k_x x) - (k_y/k_x) \sin(k_x x) 
\end{array} \right)
\eeq

Turning to the potential in Fig.~\ref{fig:squarewell}, the solution with wavevector $k_y$, has a negative (positive) energy solution in the region with $V(x)=V$ ($V(x)=-V$). The corresponding values of $k_x$ in the two regions are $k_x = ( (V-E)^2 - k_y^2)^{1/2}$ and $k_x = ( (V+E)^2 - k_y^2)^{1/2}$. So the complete transfer matrix across one period of the potential is
\beq
\mathcal{T} = \mathcal{T}_+ \left( \pi /k , ( (V+E)^2 - k_y^2)^{1/2},k_y \right)\mathcal{T}_- \left( \pi /k , ( (V-E)^2 - k_y^2)^{1/2},k_y \right).
\eeq
 Finally, by Bloch's theorem, this transfer matrix can only be a phase factor $\mathrm{e}^{\mathrm{i} 2 \pi k_x /k}$,
 where now $k_x$ is the wavevector of the Bloch eigenstates along the $x$ direction.
So we obtain the eigenvalue condition, which is $\mbox{det} (\mathcal{T} - \mathrm{e}^{\mathrm{i} 2 \pi k_x /k} ) = 0$.
Evaluating this determinant, we find the condition for a zero energy eigenvalue, $E=0$ at $k_x=0$ is
\beq
\sin (\pi (V^2 - k_y^2)^{1/2} /k) = 0,
\eeq
which leads immediately to Eq.~(\ref{qD}) for the Dirac nodes. For small $k_x$ and $E$, and with $k_y = \sqrt{V^2 - n^2 k^2} + \delta k_y$, the condition becomes
\beq
E^2 = \left( \frac{n k}{V} \right)^4 k_x^2 + \left(1 - \frac{n^2 k^2}{V^2} \right)^2 \left(\delta k_y\right)^2
\eeq
which yields the velocities in Eq.~(\ref{velocities}).

\end{appendix}

\bibliographystyle{plain}

\end{document}